\documentclass[aps,showpacs,twocolumn,superscriptaddress,prb,floatfix]{revtex4-1}
\usepackage{graphicx,color,epsfig,rotate}
\usepackage{amssymb,amsmath,bm,ulem}
\usepackage{dcolumn}  
\usepackage{hyperref} 
\usepackage[usenames,dvipsnames]{xcolor}

\newcommand{\g} \textbf{}

\def\eg{{e.g.}}

\def\epl{ Europhys.\ Lett.\ }

\def\jpcm{ J.\ Phys.:\ Condens.\ Matter }

\def\jpsj{ J.\ Phys.\ Soc.\ Jpn.\ }

\def\prb{ Phys.\ Rev.\ B }

\def\prl{ Phys.\ Rev.\ Lett.\ }

\def\rmp{ Rev.\ Mod.\ Phys.\ }

\begin{document}
\title{Dynamical structure factor of triangular-lattice antiferromagnet}
\author{M. Mourigal}
\affiliation{Institute for Quantum Matter and Department of Physics and
  Astronomy,Johns Hopkins University, Baltimore, MD 21218, USA}
\author{W. Fuhrman}
\affiliation{Institute for Quantum Matter and Department of Physics and 	
	Astronomy, Johns Hopkins University, Baltimore, MD 21218, USA}	
\author{A. L. Chernyshev}
	\affiliation{Department of Physics and Astronomy, University of California, Irvine,
	California 92697, USA}
\author{M. E. Zhitomirsky}
	\affiliation{Service de Physique Statistique, Magn\'etisme et
	Supraconductivit\'e, UMR-E9001 CEA-INAC/UJF, 17 rue des Martyrs,
	38054 Grenoble Cedex 9, France}
\date{March 16, 2016}

\begin{abstract}
We elucidate the role of magnon interaction and spontaneous decays in the spin dynamics
of the triangular-lattice Heisenberg antiferromagnet by calculating its dynamical structure
factor within the spin-wave theory.  Explicit theoretical results for neutron-scattering intensity
are provided for spins $S$ = 1/2 and $S$ = 3/2.  The dynamical structure factor exhibits
unconventional features such as quasiparticle peaks  broadened by decays,  non-Lorentzian
lineshapes, and significant spectral weight redistribution to the two-magnon continuum.
This rich excitation spectrum illustrates the complexity of the triangular-lattice antiferromagnet
and provides distinctive qualitative and quantitative fingerprints for experimental observation
of decay-induced magnon dynamics.
\end{abstract}

\maketitle

\section{Introduction}

The Heisenberg triangular-lattice antiferromagnet (HTAF) is a prominent model in low-dimensional and
frustrated magnetism and is the subject of significant experimental and theoretical interest. In zero field,
the ground-state of the model is the well-known coplanar 120$^{\circ}$ magnetic structure for all values
of spin $S$, including $S=1/2$, as evidenced by various analytical \cite{Oguchi83,Jolicoeur89,Miyake94,Chubukov94}
and numerical works. \cite{Capriotti99,Zheng06,White07} This non-collinear magnetic order has profound consequences for the spin dynamics of the HTAF: its elementary excitations (magnons)  become unstable
with respect to spontaneous decay into pairs of other magnons. \cite{Zhitomirsky13} To describe the
excitation spectrum within the spin-wave theory (SWT) the inclusion of magnon interaction is crucial. \cite{Chernyshev06,Starykh06,Chernyshev09}

Currently, a large number of materials are proposed to be fair realizations of the HTAF, although they
often deviate from the ideal model due to distorted geometry of exchange bonds or additional
spin-anisotropy terms. These include the spin-1/2 materials  $\rm Cs_2CuCl_4$, \cite{Coldea03}
$\rm Cs_2CuBr_4$, \cite{Ono03} and Ba$_3$CoSb$_2$O$_9$, \cite{Shirata12,Zhou12} and a number of compounds
with larger spin values such as VCl$_2$, \cite{Kadowaki87} LuMnO$_3$, \cite{Lewtas10} $\rm Rb_4Mn(MoO_4)_3$,
\cite{Ishii11} CuCrO$_2$, \cite{Poeinar10}, $\alpha$-SrCr$_2$O$_4$, \cite{Dutton11} and $\alpha$-CaCr$_2$O$_4$.
\cite{Toth11,Toth12}  The most comprehensive experimental characterization of these materials is done by inelastic
neutron scattering on single crystals, \cite{Poeinar10,Toth12} which directly measures the energy and momentum
dependence of spin-spin correlations as described by the dynamical structure factor $S({\bf q},\omega)$.

The spin-wave calculation of $S({\bf q},\omega)$ in the HTAF is complicated by the non-collinear spin arrangement and strong magnon interaction. Previously, the dynamical structure factor for a quasi one-dimensional spiral antiferromagnet was calculated by Ohyama and Shiba. \cite{Ohyama93} Their method was subsequently adapted to describe neutron-scattering experiments on the orthorhombically distorted triangular-lattice antiferromagnet $\rm Cs_2CuCl_4$. \cite{Veillette05,Dalidovich06} The deficiency of that method is that it operates directly with the bare Holstein-Primakoff bosons rather than with Bogolyubov quasiparticles, making the systematic $1/S$-ranking of different terms difficult and providing results that are unnecessarily complicated compared to collinear antiferromagnets. In addition, Ref.~\onlinecite{Ohyama93} does not draw a distinction between retarded and causal spin Green's functions, which is important for recovering the correct $\omega\rightarrow 0$ behavior.

One of the goals of the present work is to revisit calculation of the dynamical structure factor for a non-collinear antiferromagnet, focusing on the $1/S$-ranking  and on the correct $\omega$-dependence of various contributions to spin correlation functions. Our second goal is to provide the first explicit theoretical results for $S({\bf q},\omega)$ of the HTAF for representative values of spin to guide experimental inelastic neutron-scattering measurements in realistic materials. Such a reference point should allow evaluation of the accuracy and limits of the spin-wave theory in various experimental situations and help to identify when the latter breaks down in favor of alternative descriptions, for instance using spinons. \cite{Chung03,Kohno07,Mezio11}

The $1/S$ formalism for interacting spin-waves in the HTAF was previously described in detail in Ref.~\onlinecite{Chernyshev09}. That work focused on the role of decays in the magnon spectrum and on a classification of singularities appearing in the latter. The present work is concerned with the explicit calculation of the dynamical structure factor for the HTAF within the framework of nonlinear spin-wave theory.

Section~\ref{sec:DC} contains details of the theoretical formalism were we use basic notations from Ref.~\onlinecite{Chernyshev09}. Then, in Section~\ref{sec:Res}, we use the developed formalism  and present high-resolution predictions for the dynamical structure factor for  $S=1/2$ and $S=3/2$ along the high-symmetry directions of the Brillouin zone. Our results show a complex excitation spectrum and provide evidence for the crucial effects of magnon-magnon interactions on the spin dynamics, demonstrated by broadened quasiparticle lineshapes, double-peak structures, and contributions from the two-particle continuum that dominate a large fraction of the spectrum. 	We also present the momentum-integrated structure factor and representative constant-$\omega$ scans of $S({\bf q},\omega)$ and discuss their features. We conclude in Sec.~\ref{sec:con} and provide various details in Appendix \ref{app:A}.

\section{dynamical Correlations}
\label{sec:DC}

Neutron scattering experiments provide a direct probe of the spin-spin correlation function, otherwise known as the dynamical structure factor:
\begin{equation}
 {\cal S}^{\alpha_0\beta_0}({\bf q},\omega) = \int_{-\infty}^{\infty}\frac{dt}{2\pi}\,e^{i\omega t}
 \left\langle S^{\alpha_0}_{\bf q}(t)S^{\beta_0}_{\bf -q}(0)\right\rangle \ ,
\label{DSF}
\end{equation}
where $\alpha_0,\beta_0$ refer to spin components in the laboratory frame $\{x_0,y_0,z_0\}$. The inelastic neutron-scattering cross-section is proportional to a linear combination of the diagonal components of the correlation function (\ref{DSF}) with momentum-dependent prefactors.\cite{Squires} In the following, we do not assume a particular experimental geometry and consider instead the ``total'' structure factor in which all three components are contributing equally:
\begin{eqnarray}
{\cal S}^{\rm tot}({\bf q},\omega) = {\cal S}^{x_0x_0}({\bf q},\omega)
	+ {\cal S}^{y_0y_0}({\bf q},\omega) + {\cal S}^{z_0z_0}({\bf q},\omega).\ \ \
\label{Stot}
\end{eqnarray}

\begin{figure}[t]
\includegraphics[width = 0.999999\columnwidth]{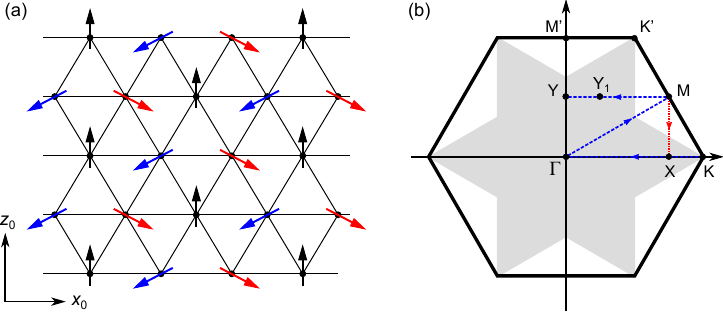}
\caption{(Color online) (a) Coplanar $120^{\circ}$ spin structure of the Heisenberg triangular-lattice antiferromagnet. (b) Brillouin zone of the triangular lattice with high-symmetry points and paths. Shaded area is the magnon decay region.}
\label{fig1}
\end{figure}

While the dynamical structure factor is measured in the laboratory reference frame, the spin-wave calculations are performed in the rotating frame with $z$ oriented along the local magnetization on each site. Using the propagation vector ${\bf Q}=(4\pi/3,0)$ of the $120^{\circ}$ spin structure, Fig.~\ref{fig1}(a), and choosing spins to rotate in the $x_0$--$z_0$ plane, the relation between spin components in the two frames is $S_i^{y_0}=S_i^{y}$ and
\begin{eqnarray}
 S_i^{x_0} & = & S_i^z \sin({\bf Q}\cdot{\bf r}_i)  + S_i^x \cos({\bf Q}\cdot{\bf r}_i)  \ ,
\nonumber \\
 S_i^{z_0} & = & S_i^z \cos({\bf Q}\cdot{\bf r}_i)  - S_i^x \sin({\bf Q}\cdot{\bf r}_i)   \ .
\label{Frametransformation}
\end{eqnarray}
Then, components of the dynamical structure factor in the two coordinate systems are connected by
\begin{eqnarray}
&& {\cal S}^{x_0x_0}({\bf q},\omega) =
 \frac{1}{4} \Bigl({\cal S}^{xx}_{{\bf q_-},\omega} +
{\cal S}^{xx}_{{\bf q_+},\omega}
+{\cal S}^{zz}_{{\bf q_-},\omega} +  {\cal S}^{zz}_{{\bf q_+},\omega}\Bigr)
\nonumber\\
&& \phantom{{\cal S}^{x_0x_0}({\bf q},\omega)}
+  \frac{i}{4}\Bigl( {\cal S}^{xz}_{{\bf q_-},\omega} - {\cal S}^{zx}_{{\bf q_-},\omega}
-{\cal S}^{xz}_{{\bf q_+},\omega} + {\cal S}^{zx}_{{\bf q_+},\omega}\Bigr)\,,
\label{Sqw_lab} \\[0.5mm]
&& {\cal S}^{z_0z_0}({\bf q},\omega) = {\cal S}^{x_0x_0}({\bf q},\omega)\,, \ \
{\cal S}^{y_0y_0}({\bf q},\omega) = {\cal S}^{yy}({\bf q},\omega)
\nonumber
\end{eqnarray}
with shorthand notations  ${\cal S}^{\alpha\beta}_{{\bf k},\omega}\equiv {\cal S}^{\alpha\beta}({\bf k},\omega)$ and ${\bf q}_\pm={\bf q\pm Q}$. In Eq. (\ref{Sqw_lab}) one can readily identify conventional diagonal contributions of the transverse ($xx$ and $yy$) and longitudinal ($zz$) spin fluctuations.\cite{Canali94}

In addition, the non-collinear nature of the ground state is responsible for terms with mixed transverse and longitudinal ($xz$ and $zx$) fluctuations. The frequency-dependence and the magnitude of the these off-diagonal correlation functions are discussed in Appendix~\ref{app:A}. We find that these off-diagonal components primarily contribute to the singularities within the two-magnon continuum that are already present in the diagonal terms, while the dominant features of the structure factor, arising from the diagonal terms, remain intact. We, thus, conclude that the off-diagonal terms always yield a sub-leading contribution with respect to the diagonal parts.  A similar conclusion was reached in the previous work on the distorted HTAF.\cite{Dalidovich06} This allows us to neglect such terms in the following consideration.

Using (\ref{Sqw_lab}), we rewrite the total structure factor in terms of the diagonal and mixed parts and further separate the former term into transverse and longitudinal contributions
\begin{eqnarray}
{\cal S}^{\rm tot}({\bf q},\omega) &=& {\cal S}^{\rm diag}({\bf q},\omega) +
  {\cal S}^{\rm mix}({\bf q},\omega) \ , \nonumber \\
{\cal S}^{\rm diag}({\bf q},\omega)
&=&  {\cal S}^{\perp}({\bf q},\omega) + {\cal S}^{L}({\bf q},\omega) \ ,   \nonumber  \\
{\cal S}^{\perp}({\bf q},\omega) &=&  {\cal S}^{yy}_{{\bf q},\omega} +
 \frac{1}{2}\left( {\cal S}^{xx}_{{\bf q_+},\omega} + {\cal S}^{xx}_{{\bf q_-},\omega} \right) ,
 \label{Stot_1} \\
{\cal S}^{L}({\bf q},\omega) &=& \frac{1}{2}
\left( {\cal S}^{zz}_{{\bf q_+},\omega} + {\cal S}^{zz}_{{\bf q_-},\omega}\right) ,	\nonumber \\
{\cal S}^{\rm mix}({\bf q},\omega)
&=&  \frac{i}{2}\left( {\cal S}^{xz}_{{\bf q_-},\omega} - {\cal S}^{zx}_{{\bf q_-},\omega}
-{\cal S}^{xz}_{{\bf q_+},\omega} + {\cal S}^{zx}_{{\bf q_+},\omega}\right). \nonumber
\nonumber
\end{eqnarray}

As discussed, we will ignore the mixed (off-diagonal) term for the bulk of this work and use explicitly  ${\cal S}^{\rm tot}({\bf q},\omega)  \approx {\cal S}^{\rm diag}({\bf q},\omega)$ in Sec.~\ref{sec:DC} and Sec.~\ref{sec:Res}. However, in Appendix~\ref{app:A} we consider the exact definition of ${\cal S}^{\rm tot}({\bf q},\omega)$ from Eq.~(\ref{Stot_1}) to illustrate the contribution of the mixed term to the total dynamical structure factor and justify our decision to neglect it.

The dynamical spin correlator ${\cal S}^{\alpha\alpha}({\bf q},\omega)$ is related to the retarded Green's function of spin operators via the fluctuation-dissipation theorem, see, {\it e.g.}, Ref.~\onlinecite{Mahan90},
\begin{equation}
{\cal S}^{\alpha\alpha}({\bf q},\omega) = - \frac{1}{\pi}\,[1+n_B(\omega)]\,
\textrm{Im}\bigl[{\cal G}^{\alpha\alpha}_{\rm ret}({\bf q},\omega)\bigr] \ ,
\label{FDT}
\end{equation}
where $n_B(\omega)=1/(e^{\omega/T}-1)$ is the Bose distribution function. Here we are interested in $T=0$ case, for which $n_B(\omega)\equiv 0$ for $\omega>0$ and $n_B(\omega)\equiv -1$ for $\omega<0$. Hence, $S({\bf q},\omega)$ is nonzero only for positive frequencies and
\begin{equation}
{\cal S}^{\alpha\alpha}({\bf q},\omega) = - \frac{1}{\pi}\,
\textrm{Im}\bigl[{\cal G}^{\alpha\alpha}_{\rm ret}({\bf q},\omega)\bigr] \ .
\label{FDT0}
\end{equation}
At $T=0$, one  can use the causal Green's function, ${\cal G}^{\alpha\beta}({\bf q},t)=-i\langle \mathcal{T} S^\alpha_{\bf q}(t) S^\beta_{\bf -q}\rangle$, on the right-hand side of Eq.~(\ref{FDT0}) since the two Green's functions coincide for $\omega>0$. This simplifies calculations, although caution is still needed when dealing with bosonic Green's functions at negative frequencies, see Sec.~\ref{sec:TF}. In the next two  subsections we consider  transverse and longitudinal components of the structure factor.

\subsection{Transverse fluctuations}
\label{sec:TF}

The spin-wave calculation of the dynamical correlation functions proceeds with the Holstein-Primakoff representation of spin operators $S_i^\alpha$ in terms of bosons $a_i$  and subsequent expansion of square roots in boson density $a_i^\dagger a_i$, see Ref.~\onlinecite{Chernyshev09} for details on application of the  SWT to the HTAF. In order to determine the leading contributions of order $O(1)$ and  $O(1/S)$ to transverse structure factor in (\ref{Stot_1}) one may use the following expressions:
\begin{equation}
 S_i^x =  \sqrt{\frac{S}{2}}\,(a_i+a_i^\dagger) \Lambda_+ \,,\ \ \
 S_i^y =-i\sqrt{ \frac{S}{2}}\,(a_i-a_i^\dagger)\Lambda_- \,,
 \label{Sxy}
\end{equation}
where
\begin{equation}
\Lambda_\pm = 1 - \frac{2n\pm\delta}{4S}
\label{Lambda}
\end{equation}
are the Hartree-Fock factors obtained from the contraction of boson operators in cubic terms with the expectation values $n = \langle a_i^\dagger a_i \rangle$ and $\delta=\langle a_i a_i \rangle$. Terms beyond the Hartree-Fock approximation contribute to the transverse structure factor in the $O(1/S^2)$ order and can be neglected.

Substituting (\ref{Sxy}) into the  spin Green's functions and performing the Bogolyubov transformation \cite{Chernyshev09} we obtain
\begin{eqnarray}
{\cal G}^{xx}({\bf q},\omega) & = & \frac{S}{2}\,\Lambda_+^2(u_{\bf q}+v_{\bf q})^2
 \bigl[G_{11}({\bf q},\omega)\!
 \nonumber \\
 &  & \phantom{\frac{S}{2}} + G_{11}(-{\bf q}, -\omega) + 2 G_{12}({\bf q}, \omega) \bigr],
 \nonumber \\
{\cal G}^{yy}({\bf q},\omega) & = & \frac{S}{2}\,\Lambda_-^2(u_{\bf q}-v_{\bf q})^2
 \bigl[G_{11}({\bf q},\omega)
\label{G_trans} \\
 &  & \phantom{\frac{S}{2}} + G_{11}(-{\bf q},-\omega) - 2 G_{12}({\bf q}, \omega)\bigr].
 \nonumber
 \end{eqnarray}
Here $G_{11}({\bf q},\omega)$ and $G_{12}({\bf q},\omega)$ are the normal and anomalous magnon Green's functions and  $u_{\bf q}$ and $v_{\bf q}$ are the Bogolyubov coefficients.

In the harmonic approximation, $G_{12}({\bf q},\omega) \equiv 0$ and $G_{11}({\bf q},\omega) \equiv  G_0({\bf q},\omega)= (\omega - \varepsilon_{\bf q} + i0)^{-1}$, where $\varepsilon_{\bf q}$ is the magnon energy in the harmonic approximation
\begin{equation}
\varepsilon_{\bf q} = 3JS\sqrt{(1-\gamma_{\bf q})(1+2\gamma_{\bf q})} \ ,
\end{equation}
with $\gamma_{\bf q} = \frac{1}{3}\big[\cos q_x +2 \cos(\frac{q_x}{2})\cos(\frac{\sqrt{3}}{2} q_y ) \big]$. Hence, in this approximation, magnon excitations manifest themselves as sharp delta-peaks in the dynamical structure factor. However, in spiral antiferromagnets, magnon-magnon interaction alters substantially this simplified picture.  The complication is mainly due to three-magnon processes, which are inherent to noncollinear antiferromagnets, \cite{Zhitomirsky13} and produce the $\omega$-dependent self-energy already in the lowest-order perturbation theory, also leading to a finite lifetime of magnons in a large part of the Brillouin zone.

For the purpose of the $1/S$-ranking of various contributions we note that the magnon energy scales with spin as $\varepsilon_{\bf q} = O(S)$ and the self-energy as $\Sigma_{11,12}({\bf q},\omega) = O(1)$. Then,  to achieve the $O(1/S)$ accuracy in the structure factor, one can use a reduced form of the Belyaev equations for the magnon Green's functions:
\begin{eqnarray}
 G_{11}({\bf q},\omega) & \approx & 1/\bigl[\omega-\varepsilon_{\bf q}-\Sigma_{11}({\bf q},\omega)\bigr] \ ,
 \nonumber \\[0.5mm]
 G_{12}({\bf q},\omega) & \approx & \Sigma_{12}({\bf q},\omega)G_{11}({\bf q},\omega)G_{11}(-{\bf q},-\omega)\ .
 \label{BE}
\end{eqnarray}
Clearly, $G_{11}({\bf q},\omega)=O(1/S)$ and  $G_{12}({\bf q},\omega)=O(1/S^2)$.
In the lowest  order, magnon self-energies are
\begin{eqnarray}
\Sigma_{11}({\bf q},\omega) & = &  \Sigma_{11}^{\rm HF}({\bf q}) + \frac{1}{2} \sum_{\bf k}
\frac{|V_{31}({\bf k};{\bf q})|^2}{\omega-\varepsilon_{\bf k}-\varepsilon_{\bf q-k}+i0}
\nonumber  \\
&&\phantom{\Sigma_{11}^{\rm HF}({\bf q})} - \frac{1}{2}\sum_{\bf k} \frac{|V_{32}({\bf k},{\bf q})|^2}
{\omega + \varepsilon_{\bf k} + \varepsilon_{\bf q+k} \mp i0}, \nonumber   \\
\Sigma_{12}({\bf q},\omega)  &=&  \Sigma_{12}^{\rm HF}({\bf q}) + \frac{1}{2}\sum_{\bf k}
\frac{V_{32}({\bf k},-{\bf q})V^*_{31}({\bf k};{\bf q})}
{\omega - \varepsilon_{\bf k} - \varepsilon_{\bf q-k}+i0}  \ \ \ \ \
\label{Sigma}\\
&& \phantom{\Sigma_{11}^{\rm HF}({\bf q})} - \frac{1}{2}\sum_{\bf k}\frac{V_{32}({\bf k},{\bf q})V^*_{31}({\bf k};-{\bf q})}
{\omega  + \varepsilon_{\bf k} + \varepsilon_{\bf q+k} \mp i0},  	\nonumber
\end{eqnarray}
where $\Sigma^{\rm HF}({\bf q})$ are the frequency-independent Hartree-Fock contributions, $V_{31}({\bf k};{\bf q})$ and $V_{32}({\bf k},{\bf q})$ are the three-particle  decay and source  interaction vertices, respectively, and $\mp i0$ correspond to the causal/retarded self-energies. \cite{Chernyshev09} We note that one must  use the lower sign  in (\ref{Sigma}) to ensure the correct odd-frequency dependence of  the imaginary part of the magnetic susceptibility. In the following, we use small  $\delta \equiv 0^+$ for the numerical evaluation of the self-energies in (\ref{Sigma}).

Several important simplifications are in order. The  term containing anomalous Green's function on the right-hand sides of Eqs.~(\ref{G_trans}) is next order in $1/S$ classification compared to the first two. While it does contribute to the structure factor in the sought $O(1/S)$ order, its contribution can be shown to be small already for $S=1/2$ and also qualitatively redundant to that of the other terms, see Appendix~\ref{app:A} for analysis. We therefore neglect these terms in the following consideration.

Formally, the first two  Green's functions on the right-hand side of Eqs.~(\ref{G_trans}) are of the same order in the $1/S$-ranking and could contribute equally to $\textrm{Im}\bigl[{\cal G}^{\alpha\alpha}({\bf q},\omega)\bigr]$. However, the second term, $\textrm{Im}\bigl[G_{11}({\bf -q},-\omega)\bigr]$, is off-resonance compared to $\textrm{Im}\bigl[G_{11}({\bf q},\omega)\bigr]$ and contains no poles for $\omega>0$, thus providing no contribution to the structure factor. While this term is important to ensure the correct behavior  of the spectral function for low-energy excitations at $\omega, \varepsilon_{\bf q}\rightarrow 0$,\cite{Talbot88} for all practical purposes it is negligible.

One should note that the consideration of spectral properties within the SWT always exceeds the nominal $1/S$ order as the frequency-dependence is automatically ``off-shell'', thus including contributions of higher $1/S$ order. A technical issue arises when evaluating magnon spectral function with $\Sigma_{11}({\bf q},\omega)$ given by Eq.~\eqref{Sigma}. Due to the third term in \eqref{Sigma} (``source"), a spurious excitation branch appears in the vicinity of ${\bf Q}$ with vanishing $\varepsilon_{\bf q}$ at a ${\bf q}\neq {\bf Q}$. This unphysical mode is manifestation of a pole pushed up from negative values of $\omega$. Similar behavior is also present in $\textrm{Im}\bigl[G_{11}({\bf -q},-\omega)\bigr]$, which develops an extension of the same mode in the formally forbidden $\omega>0$ region. In principle, these anomalies should be removed by some self-consistent higher-order $1/S$ expansion, an analytically and computationally difficult problem.  Here we adopt a more expeditious manner to address the un-physical mode directly by returning the offending source self-energy term back on-shell, i.e. taking $\omega=\varepsilon_{\bf q}$ within this term. In this way, the effect of magnon-magnon interactions and  decays are maintained while unphysical singularities are suppressed.

Altogether, for the results of the next Section we use the following expressions for the ${\cal S}^{yy}({\bf q},\omega)$ and ${\cal S}^{xx}({\bf q},\omega)$ components of the transverse structure factor in (\ref{Stot_1})
\begin{eqnarray}
{\cal S}^{xx}({\bf q},\omega) & = & \frac{S}{2}\,\Lambda_+^2(u_{\bf q}+v_{\bf q})^2
 A_{11}({\bf q},\omega) ,
 \nonumber \\
{\cal S}^{yy}({\bf q},\omega) & = & \frac{S}{2}\,\Lambda_-^2(u_{\bf q}-v_{\bf q})^2
A_{11}({\bf q},\omega),
\label{G_trans_1}
 \end{eqnarray}
with $A_{11}({\bf q},\omega)=-(1/\pi)\textrm{Im}\bigl[G_{11}({\bf q},\omega)\bigr]$, where $G_{11}({\bf q},\omega)$ is determined by Eq.~(\ref{BE}) with $\Sigma_{11}({\bf q},\omega)$ given by Eq.~\eqref{Sigma}, in which the source term is taken on-shell, $\omega=\varepsilon_{\bf q}$. We note that in the harmonic approximation
${\cal S}^{xx}[{\cal S}^{yy}]({\bf q},\omega)  \sim S A_{11}({\bf q},\omega) \sim S \delta(\omega-\varepsilon_{\bf q})$, which is explicitly of order $O(1)$.

\subsection{Longitudinal fluctuations}
\label{sec:LF}

Spin fluctuations in the direction of ordered moments written in terms of Holstein-Primakoff bosons are
\begin{equation}
\delta S_{\bf q}^z = - \frac{1}{\sqrt N}\sum_{\bf k} a_{\bf k}^\dagger a_{\bf k+q} \ .
\label{dSz}
\end{equation}
Then, the inelastic part of the longitudinal neutron cross-section ${\cal S}^{L}({\bf q},\omega)$ in (\ref{Stot_1})  is determined by the correlation function ${\cal S}^{zz}({\bf q},t)=\left\langle \delta S_{\bf q}^z(t)\delta S_{-{\bf q}}^z\right\rangle$, which probes the two-magnon continuum.

As is clear from the derivation given in Appendix~\ref{app:A}, the longitudinal component of the structure factor is of order $O(1/S)$, a factor of $1/S$ smaller than the leading terms in the transverse correlation functions (\ref{G_trans_1}). Therefore, in the spirit of the $1/S$ expansion it may be calculated with bare magnon Green's functions, neglecting  corrections from magnon-magnon interactions. Since this approximation neglects renormalization of the magnon energies, the obtained width of the two-magnon continuum will not be precise, but will still give a fair embodiment of the continuum contribution to the neutron-scattering cross-section.

With this we obtain
\begin{equation}
{\cal S}^{zz} ({\bf q},\omega)  =
\frac{1}{2}\! \sum_{\bf k} (u_{\bf k} v_{\bf k-q}\! + v_{\bf k} u_{\bf k-q})^2
\delta(\omega-\varepsilon_{\bf k} - \varepsilon_{\bf k-q}),
\label{Szzqw}
\end{equation}
see Appendix~\ref{app:A} for details.

Thus, in the following calculations of ${\cal S}^{L}({\bf q},\omega)$ in (\ref{Stot_1}) we we shall use  ${\cal S}^{zz} ({\bf q},\omega)$ from Eq.~(\ref{Szzqw}).

\begin{figure*}[t!]
\includegraphics[width=0.63\textwidth]{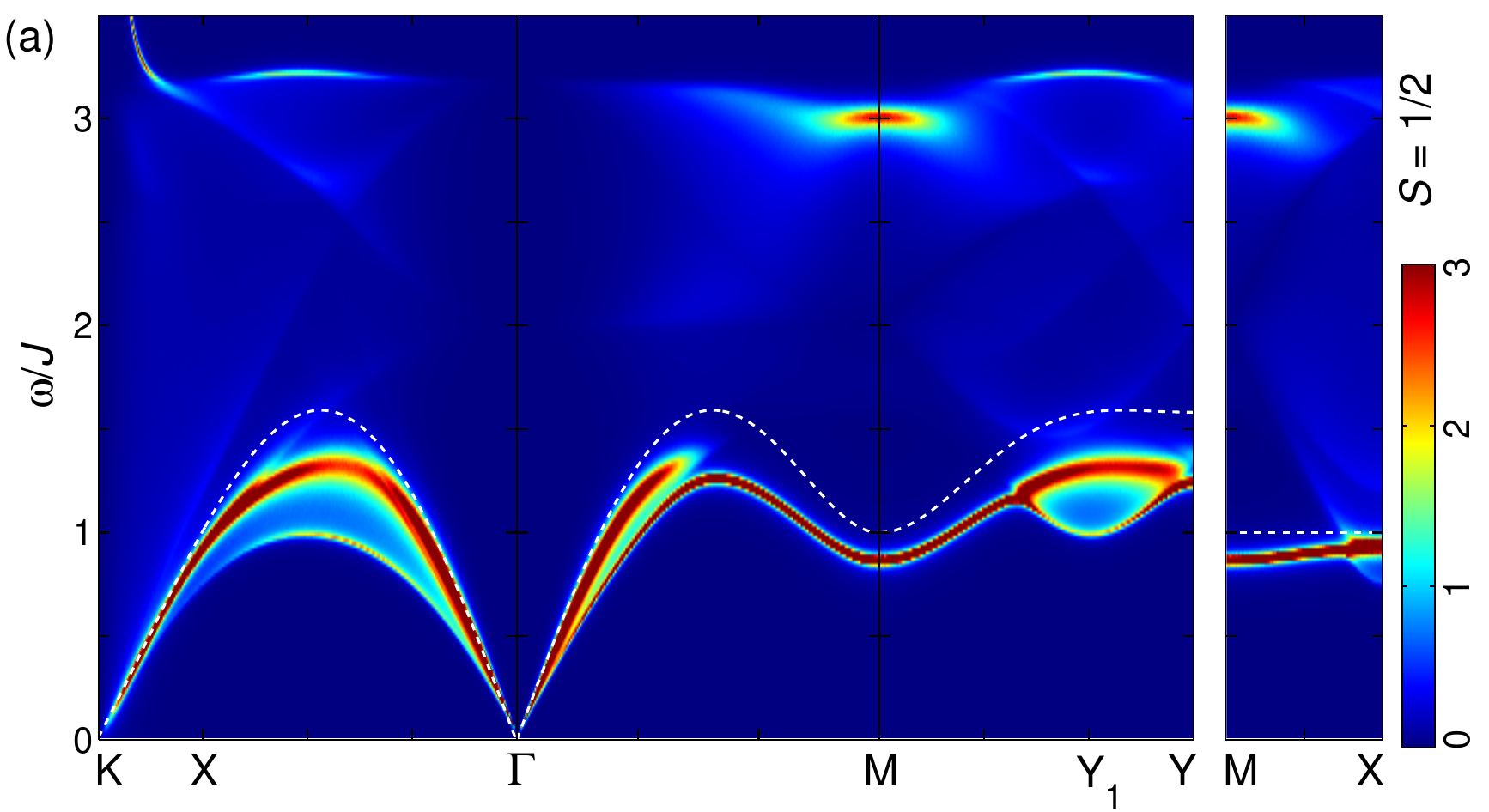}
\includegraphics[width=0.63\textwidth]{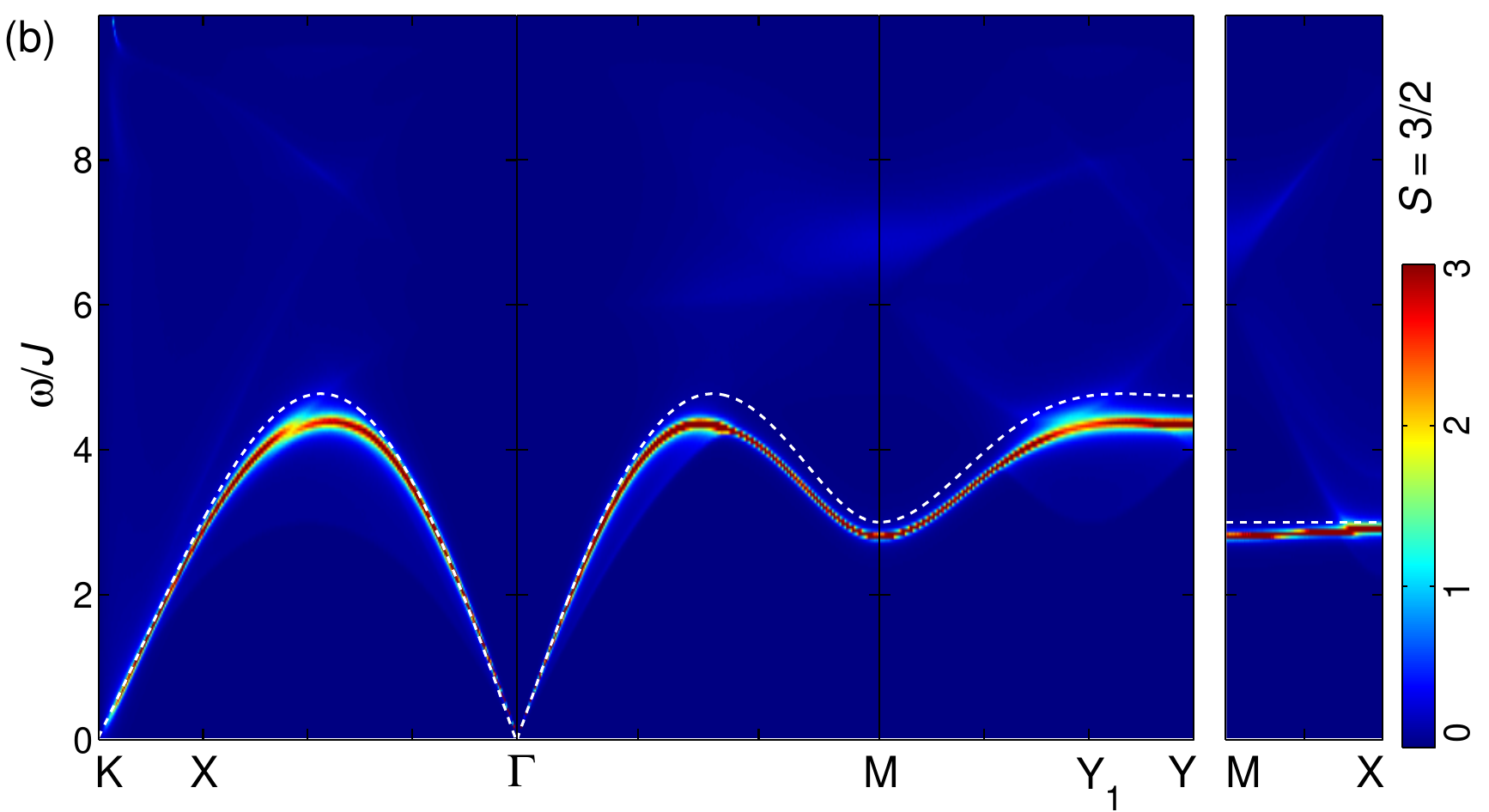}
\caption{
(Color online) Intensity plots of the momentum and energy dependence of the spectral function $A_{11}({\bf q,\omega})$  for (a) $S=1/2$ and (b) $S=3/2$ along the high-symmetry paths in the Brillouin zone in Fig.~\ref{fig1}(b). Dashed line is the linear SWT spectrum $\varepsilon_{\bf q}$.
}
\label{fig2}
\end{figure*}

\section{results and discussion}
\label{sec:Res}

In this section we present high-resolution numerical results using Eqs.~(\ref{Stot_1}), (\ref{G_trans_1}), (\ref{Szzqw}) and ${\cal S}^{\rm tot}({\bf q},\omega) \approx {\cal S}^{\rm diag}({\bf q},\omega)$ to provide specific predictions for the dynamical structure factor of the HTAF  for $S=1/2$ and $S=3/2$, revealing the dramatic effects of magnon-magnon interactions. We begin with the analysis of transverse fluctuations related to the normal part of the spectral function and proceed to the comparison of the relative weights of transverse and longitudinal fluctuations in the dynamical structure factor for representative momenta. Finally, we show our results for the total dynamical structure factors of the HTAF for $S=1/2$ and $S=3/2$ and conclude with the description of their momentum-integrated forms.

We performed the numerical integration of the self-energies in Eq.~(\ref{Sigma}) using an artificial broadening of $\delta = 0.03JS$ and various integration schemes. The intensity plots of the spectral function (Fig.~\ref{fig2}) and dynamical structure factor (Fig.~\ref{fig4}) used a quasi-Monte-Carlo integration in {\scshape Mathematica} with an accuracy goal of four digits. The line plots of Figs.~\ref{fig3} and \ref{fig7} used a Gaussian-quadrature method with $4\cdot10^6$ points while a simple Monte-Carlo integration with $5\cdot10^6$ points was used for Fig.~\ref{fig7}. The momentum-integrated ${\cal S}^{\rm tot}(\omega)$ and constant-energy ${\bf q}$-scans of ${\cal S}^{\rm tot}({\bf q},\omega)$ in Figs.~\ref{fig5}, \ref{fig6} and \ref{fig6a} used a Gaussian-quadrature method with $1.6\cdot10^5$ ${\bf k}$ and ${\bf q}$ points and a somewhat larger $\delta = 0.04JS$. A higher density mesh of $1.44\cdot10^6$ points was used in Fig.~\ref{fig5} for the long-wavelength region $\omega/SJ<0.5$ with the  subsequent extrapolation to $\omega\rightarrow0$ limit.

\subsection{Transverse fluctuations and spectral function}
\label{sec:ResTF}

We begin with the examination of the transverse dynamical structure factor ${\cal S}^{\perp}({\bf q},\omega)$ in (\ref{Stot_1}). The dominant contribution to this component originates from the normal part of the magnon Green's function $G_{11}({\bf q},\omega)$, with momenta ${\bf q}$ and ${\bf q}\pm{\bf Q}$. Neglecting the anomalous terms, ${\cal S}^{\perp} ({\bf q},\omega)$ is related to the spectral function $A_{11}({\bf q},\omega) = -1/\pi\,{\rm Im}[G_{11}({\bf q},\omega)]$ with momentum-dependent pre-factors, see Eq.~(\ref{G_trans_1}). Therefore, in Fig.~\ref{fig2} we restrict ourselves with $A_{11}({\bf q,\omega})$ for $S\!=\!1/2$ and $S\!=\!3/2$ along the high-symmetry directions in the Brillouin zone from Fig.~\ref{fig1}(b). This presentation gives the benefit of relative simplicity and highlights important features of the spectrum related to magnon interactions and decays, which will be subsequently identified in the more complicated landscape of the structure factor. The transverse dynamical structure factor is then obtained by a linear combination of $A_{11}({\bf q},\omega)$ and  $A_{11}({\bf q}\pm{\bf Q},\omega)$ according to (\ref{Stot_1}) and (\ref{G_trans_1}).

The effect of magnon interaction is taken into account by the self-energies (\ref{Sigma}), which originate from a direct coupling of the single-particle branch to the two-magnon continuum. Because of such a coupling,  an incoherent component is present in the intensity plots of $A_{11}({\bf q,\omega})$ in Fig.~\ref{fig2}, which also provides an insight into the quasiparticle-like behavior of the single-particle excitations, potentially broadened by decays.\cite{Zhitomirsky13}

A feature of the HTAF spectrum, observed for all  momenta, is the downward renormalization of the magnon dispersion from its bare value $\varepsilon_{\bf q}$ (dashed line) for both $S\!=\!1/2$ and $S\!=\!3/2$, see Fig.~\ref{fig2}. This is in agreement with a number of previous works on the HTAF\cite{Starykh06,Chernyshev06,Chernyshev09,Zheng06} and on related problems involving magnon interaction in noncollinear antiferromagnets. \cite{Mourigal10,Fuhrman12,Zhitomirsky13} This generic effect is due to an effective repulsion between the single-particle branch and the two-particle continuum facilitated by their coupling. Such renormalization is about 18\% for $S\!=\!1/2$ in Fig.~\ref{fig2}(a) and about 8\% for $S\!=\!3/2$ in Fig.~\ref{fig2}(b), representing a quantum $1/S$-effect. The renormalization factor for $S=1/2$ is somewhat less than in the numerical\cite{Zheng06} and on-shell SWT results,\cite{Starykh06,Chernyshev06,Chernyshev09} but is in closer agreement with the results of the off-shell Dyson equation SWT approach.\cite{Chernyshev09} Other aspects of the spectrum renormalization, such as development of the ``roton-like'' minimum at the M-point, are also in agreement with earlier studies.\cite{Starykh06,Chernyshev06,Zheng06} Note that the discussed effect of magnon interaction in the HTAF is in contrast with the well-known upward spectrum renormalization for the collinear antiferromagnets. \cite{Igarashi05}

An interesting  signature of strong magnon interaction in the $S=1/2$ case is the bright intensity spot at $\omega/J\approx 3$ in the vicinity of the M-point, see Fig.~\ref{fig2}(a). On a closer examination we find that this is an antibonding state: the single-magnon state pushed out of the two-magnon continuum. While this state is likely an artifact of our approximation and  will broaden  significantly if treated self-consistently, it signifies the strength of the magnon-magnon interaction. Note that this effect disappears in the $S=3/2$ case where magnon interaction is weaker.

Another prominent feature is the broadening of the quasiparticle peaks  observed for the momenta inside the decay region (shaded area of the Brillouin zone in Fig.~\ref{fig1}(a)), \eg, along the K$\Gamma$-line as well as in the $\Gamma$M and MY directions. A particularly salient broadening occurs  in $S=1/2$ case, as seen in Fig.~\ref{fig2}(a). The corresponding  magnon excitations acquire a finite lifetime due to three-particle magnon-magnon interactions. \cite{Zhitomirsky13}The kinematic conditions required for such processes are discussed in detail elsewhere,\cite{Chernyshev09} though we note that the boundary of the  decay region is due to emission of the acoustic magnon $\varepsilon_{\bf Q}$. This is distinct from the case of magnetic-field induced decays in the square-lattice antiferromagnet where the corresponding decay products are inside the decay region and thus also unstable.\cite{Zhitomirsky99,Mourigal10} As a consequence, the boundaries of the HTAF decay region are sharply defined, leading to a spectacular and robust quasiparticle ``blow-out" when the single-magnon branch enters the decay region and merges with the two-magnon continuum, as visible  along the MY-path in Fig.~\ref{fig2}(a). This effect resembles neutron scattering observations of the so-called termination point for the excitations of superfluid $^4$He\cite{Fak98} and the triplet excitations of spin-gap materials.\cite{Stone06,Masuda06} Similar  distortion of the excitation curve in the vicinity of a continuum boundary was also observed in the other spin systems.\cite{Coldea10}

\begin{figure}[b]
\includegraphics[width=0.9999\columnwidth]{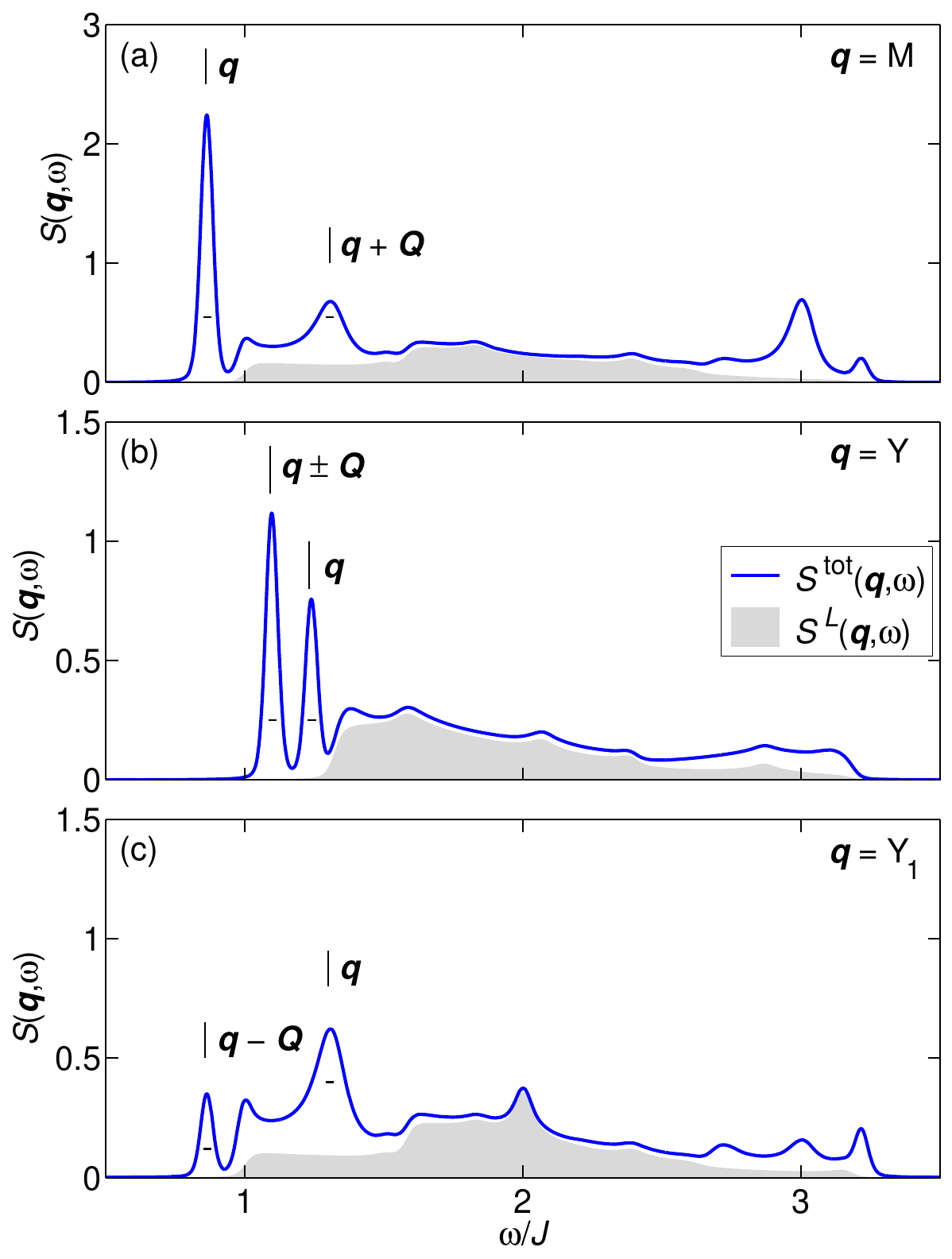}
\caption{
(Color online) (a)--(c) Energy dependence of the dynamical structure factor for $S=1/2$ at representative  points of the Brillouin zone, M, Y and Y$_1$, see Fig.~\ref{fig1}(b). Solid line corresponds to the total structure factor ${\cal S}^{\rm tot}({\bf q},\omega)$, shaded area is the longitudinal part ${\cal S}^{L}({\bf q},\omega)$. The  vertical marks indicate positions of the quasiparticle peaks from $A_{11}({\bf q},\omega)$ and  $A_{11}({\bf q}\pm{\bf Q},\omega)$. Horizontal bars indicate the width $\sigma = 0.03J$ of the Gaussian convolution.}
\label{fig3}
\end{figure}

One can see in Fig.~\ref{fig2}(a) that  the crossing between one-particle spectrum and  two-magnon continuum is accompanied by the ``edge" singularity, visible as the lowest-energy branch  for the K$\Gamma$-line or as a ``double-peak'' structure for the $\Gamma$M-path if cutting along the $\omega$ axis. Such features are the van Hove singularities\cite{Chernyshev09} due to the bottom of the two-magnon continuum, see also Figs.~\ref{fig3}, \ref{fig4} and \ref{fig7}. Within the SWT, they should be regularized by the higher-order diagrams\cite{Chernyshev09} and in realistic systems by a small inter-layer coupling.\cite{Fuhrman12}

While the role of interaction between magnons decreases for $S = 3/2$, magnon decays remain highly visible, in particular along the K$\Gamma$-line and in the ``blow-out'' region of the MY-line, see Fig~\ref{fig2}(b). The spectral weight transfer from the single-particle excitations to  the continuum is, however, strongly reduced.

Finally, we note that many of the spectral features discussed here, such as spectrum broadening throughout large regions of the Brillouin zone, dramatic redistributions of spectral weight to the two-magnon continuum, non-Lorentzian two-peak structures, and other features clearly unlike conventional single-particle peaks have been discussed by us recently for the quasi-2D square-lattice antiferromagnet in a field.\cite{Fuhrman12,Zhitomirsky13}

\subsection{Total dynamical structure factor}
\label{sec:ResDSF}

We now proceed with the analysis of the total dynamical structure factor ${\cal S}^{\rm tot}({\bf q},\omega) \approx {\cal S}^{\rm diag}({\bf q},\omega)$ in (\ref{Stot_1}) and of the role of  the longitudinal component ${\cal S}^{L}({\bf q},\omega)$ in it, which we take in the form given by Eq.~(\ref{Szzqw}). In Fig.~\ref{fig3} we offer such a consideration for $S\!=\!1/2$ and for representative high-symmetry points, M, Y and Y$_1$, see Fig.~\ref{fig1}(b). Note that Y$_1$ is the image of the M-point shifted by the ${\bf Q}$ vector.

\begin{figure*}[th!]
\includegraphics[width=0.63\textwidth]{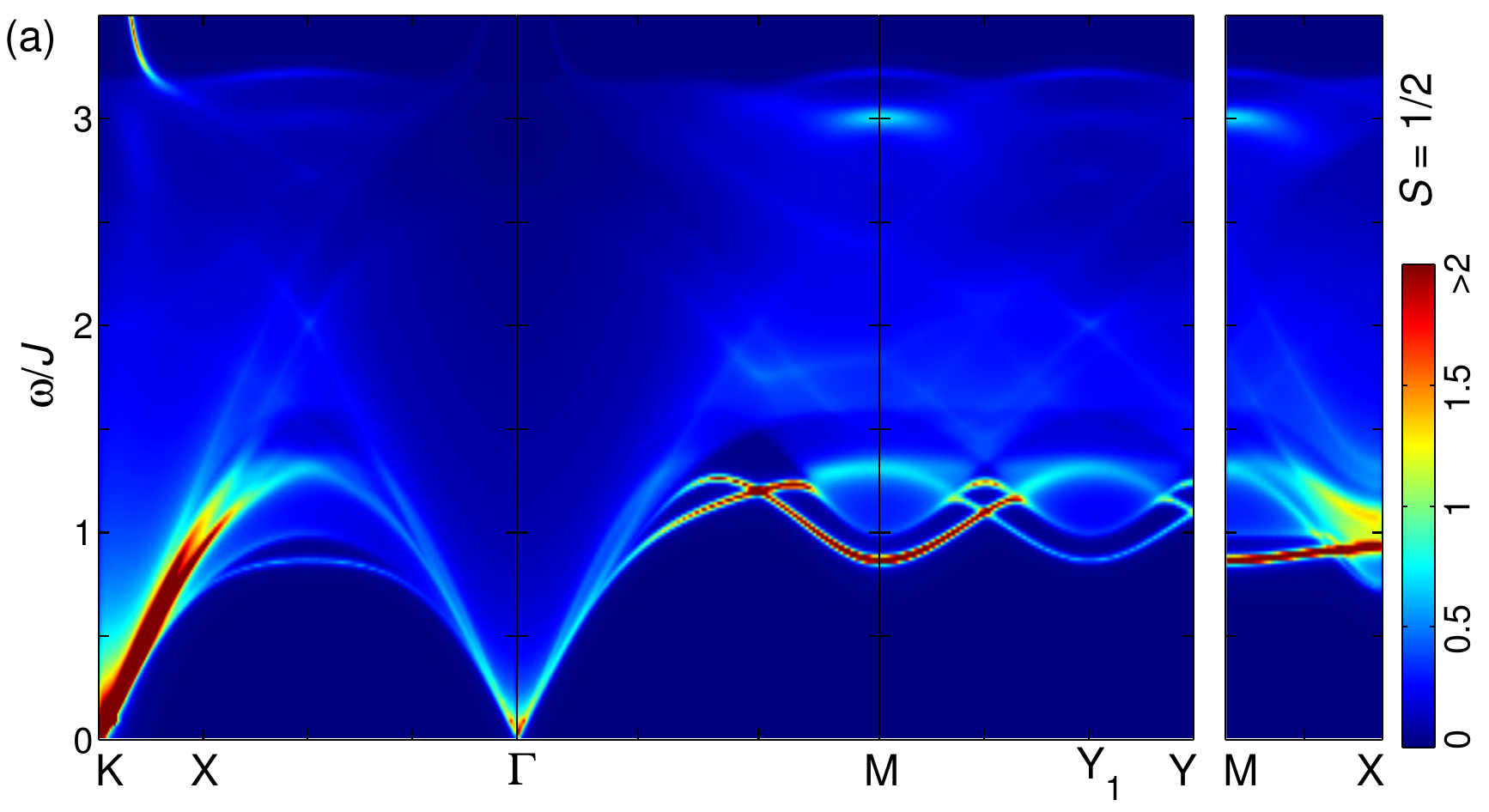}
\includegraphics[width=0.63\textwidth]{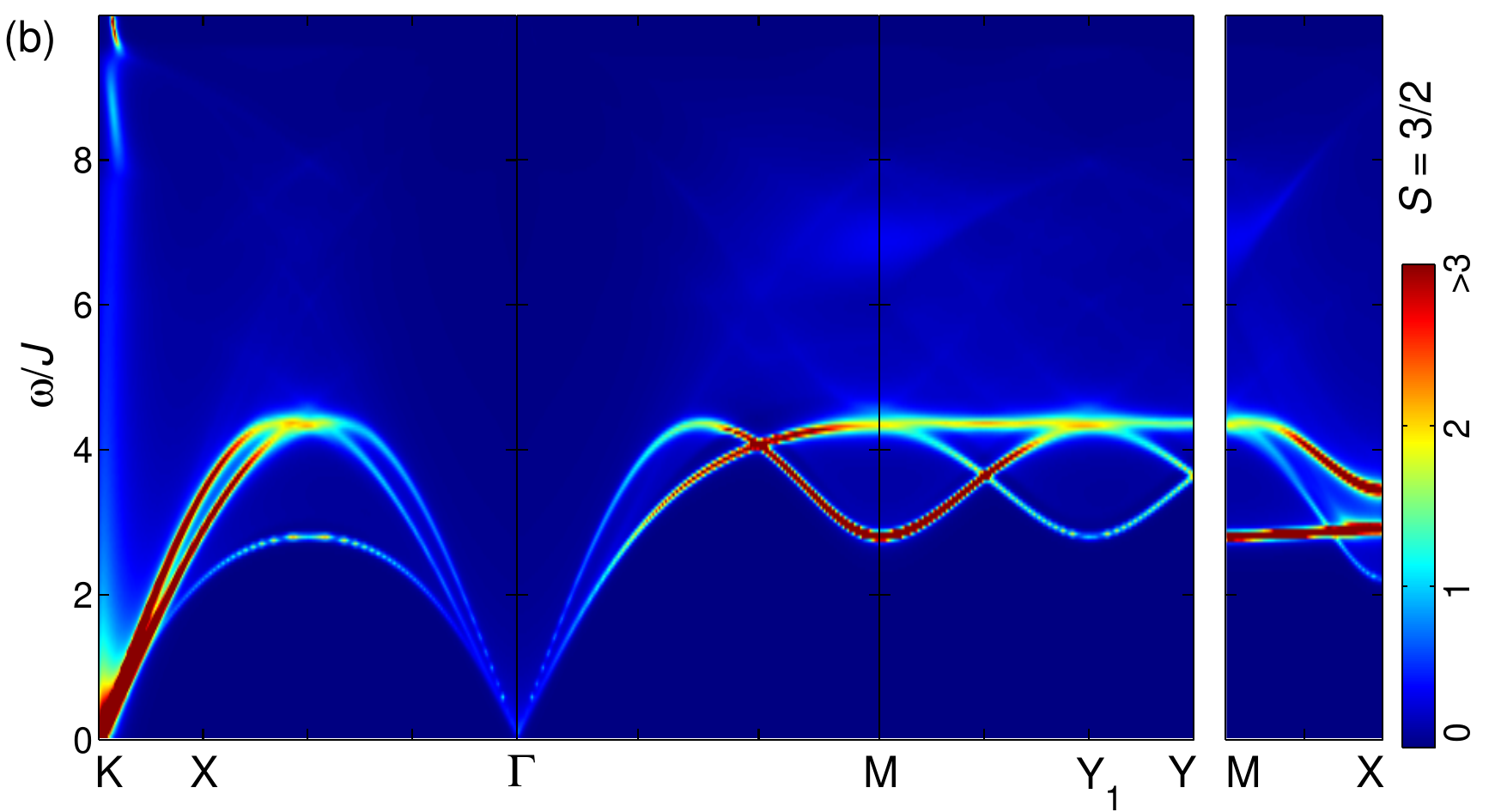}
\caption{(Color online)
Intensity plots of the momentum and energy dependence of the total dynamic structure factor $S^{\rm tot}({\bf q,\omega})$ (\ref{Stot_1}) for (a) $S=1/2$ and (b) $S=3/2$ along the high-symmetry paths in the Brillouin zone shown in Fig.~\ref{fig1}(b).}
\label{fig4}
\end{figure*}

The contribution of the longitudinal component to the total dynamical structure factor is shown in Fig.~\ref{fig3} by shaded areas while the total structure factor is plotted by solid lines. In order to mimic a hypothetical experimental energy resolution as well as to soften various spurious features such as the  edge-singularities in ${\cal S}^{\perp}({\bf q},\omega)$ discussed above or the van Hove singularities of the two-magnon density of states in ${\cal S}^{L}({\bf q},\omega)$, the  results are convoluted with a Gaussian profile of $\sigma=0.03J$.  This is indicated by  horizontal bars in Fig.~\ref{fig3} and done in addition to the artificial broadening  $\delta$ used in the numerical integration.

Several aspects of the results presented in Fig.~\ref{fig3} deserve mentioning. As we discussed above, at each ${\bf q}$-point the transverse component of the structure factor is a linear combination of three spectral functions, $A_{11}({\bf q},\omega)$, $A_{11}({\bf q}-{\bf Q},\omega)$, and  $A_{11}({\bf q}+{\bf Q},\omega)$ with different ${\bf q}$-dependent weights, see Eqs.~(\ref{G_trans_1}) and (\ref{Stot_1}). For the high-symmetry points of our choice, only two of such terms are distinct. Given the correspondence between ${\bf q}_\textrm{M}\pm{\bf Q}$ and the points equivalent to Y$_1$,  the similarity of their structure factors, positions and shapes of the peaks, and other features in Figs.~\ref{fig3}(a) and (c) are now easily understood. One can also see that the ${\bf q}$-dependent weights yield different relative intensity of different features at M and Y$_1$ points. Moreover, using our previous analysis of the spectral function one can observe that the lowest peaks in Figs.~\ref{fig3}(a) and (c) are resolution-limited and both come from the sharply-defined peak in $A_{11}({\bf q},\omega)$ at the M-point in Fig.~\ref{fig2}, which is outside the decay region. At the same time,  quasiparticle peaks that are broadened by decays and accompanied by the non-Lorentzian edge-like features below them, marked  as ${\bf q}+{\bf Q}$ and ${\bf q}$ in Figs.~\ref{fig3}(a) and (c), respectively, originate from the same spectral shapes in $A_{11}({\bf q},\omega)$ at the Y$_1$-point, see Fig.~\ref{fig2}.

For the structure factor at the Y-point, the transverse part is dominated by the two well-defined quasiparticle peaks. While the lowest one [${\bf q}\pm{\bf Q}$] corresponds to stable magnons, the second peak is from the Y-point itself, which is inside the decay region in Fig.~\ref{fig1}(b).  A closer inspection of Fig.~\ref{fig2} and the on-shell analysis in Ref.~\onlinecite{Chernyshev09} show that the corresponding broadening due to decays is small for this point.  A similar type of comprehensive analysis of the structure factor is possible for any other ${\bf q}$-point.

A compelling property of the structure factor for $S\!=\!1/2$ in Fig.~\ref{fig3} is a very strong contribution of the longitudinal component ${\cal S}^{L}({\bf q},\omega)$ for each momenta.  This is directly related to the transfer of part of the static spectral weight (reduction of the ordered moment) to the longitudinal dynamical correlations under the action of strong quantum fluctuations. Thus, in addition to the broad, continuum-like features of the transverse structure factor, the longitudinal component dominates  the wide range of $\omega$ in each of the plots in Fig.~\ref{fig3}. In fact, it contributes the major portion to the total spectral weight at the Y-point in  Fig.~\ref{fig3}(b).

To complete the discussion of Fig.~\ref{fig3}, we also note that while the transverse parts of the structure factors at the M and Y$_1$ points are related, the corresponding longitudinal components are different. This is because, according to Eq.~(\ref{Stot_1}), ${\cal S}^{L}({\bf q},\omega)$ at a ${\bf q}$-point takes two contributions, from ${\cal S}^{zz}({\bf q+Q},\omega)$ and  ${\cal S}^{zz}({\bf q-Q},\omega)$. Hence, given the relation between  M and Y$_1$, $S^L({\bf q}_\textrm{M},\omega) = {\cal S}^{zz}({\bf q}_{\rm Y_1},\omega)$, but not vice versa.

In Fig.~\ref{fig4} we present the intensity plots of the ${\bf q}$- and $\omega$-dependence of the total structure factor for both $S=1/2$ and $S=3/2$ along the high-symmetry paths in Fig.~\ref{fig1}(b). One of the benefits of the insight provided by our preceding discussion of the spectral function in Fig.~\ref{fig2} and of ${\cal S}({\bf q},\omega)$ at selected ${\bf q}$-points in Fig.~\ref{fig3} is that now the complicated view of Fig.~\ref{fig4} is seen naturally as a superposition of three ${\bf q}$-modulated $A_{11}({\bf q},\omega)$ terms and a background of two ${\cal S}^{zz}({\bf q},\omega)$ terms.

As can be anticipated from the richness in the behavior of the spectral function, the total dynamical structure factor ${\cal S}^{\rm tot}({\bf q},\omega)$ shows a  complex interplay of quasiparticle-like and continuum contributions, revealing an abundant  broadening of the peaks coexisting with the sharply-defined excitations that are brought in by the shifted $\pm {\bf Q}$ branches. For instance, the spectacular ``blow-out'' region of the single-magnon branch entering the two-magnon continuum along the YM direction near Y$_1$ in Fig.~\ref{fig2}(a) now acquires a ``mirror'' region around the M-point. Note that  both of these also coexist with the well-defined magnon branches at lower energy.

In addition, for $S=1/2$, the continuum-like component dominates the spectrum  throughout the Brillouin zone at the higher energies. For the $S=3/2$ case in Fig.~\ref{fig4}(b), the full structure factor is composed of three well-defined quasiparticle branches, which also demonstrate  some substantial continuum-like scattering. Altogether, the total structure factor exhibits a complex landscape consisting of sharp and broadened quasiparticle peaks as well as substantial continuum contributions  from both the transverse and the longitudinal parts of the dynamical structure factor.

In order to analyze the contributions of the continuum and of the quasiparticle-like excitations
to the structure factor on a more quantitative level we consider the {\it momentum-integrated} dynamical structure factor
\begin{equation}
{\cal S}^{\rm tot}(\omega) = \sum_{\bf q}  {\cal S}^{\rm tot}({\bf q},\omega) \ ,
\end{equation}
which coincides with the spectral density of the spin autocorrelation function. Such a quantity is readily accessible in neutron-scattering experiments on powder samples. \cite{Toth12}

\begin{figure}[t!]
\includegraphics[width=0.99\columnwidth]{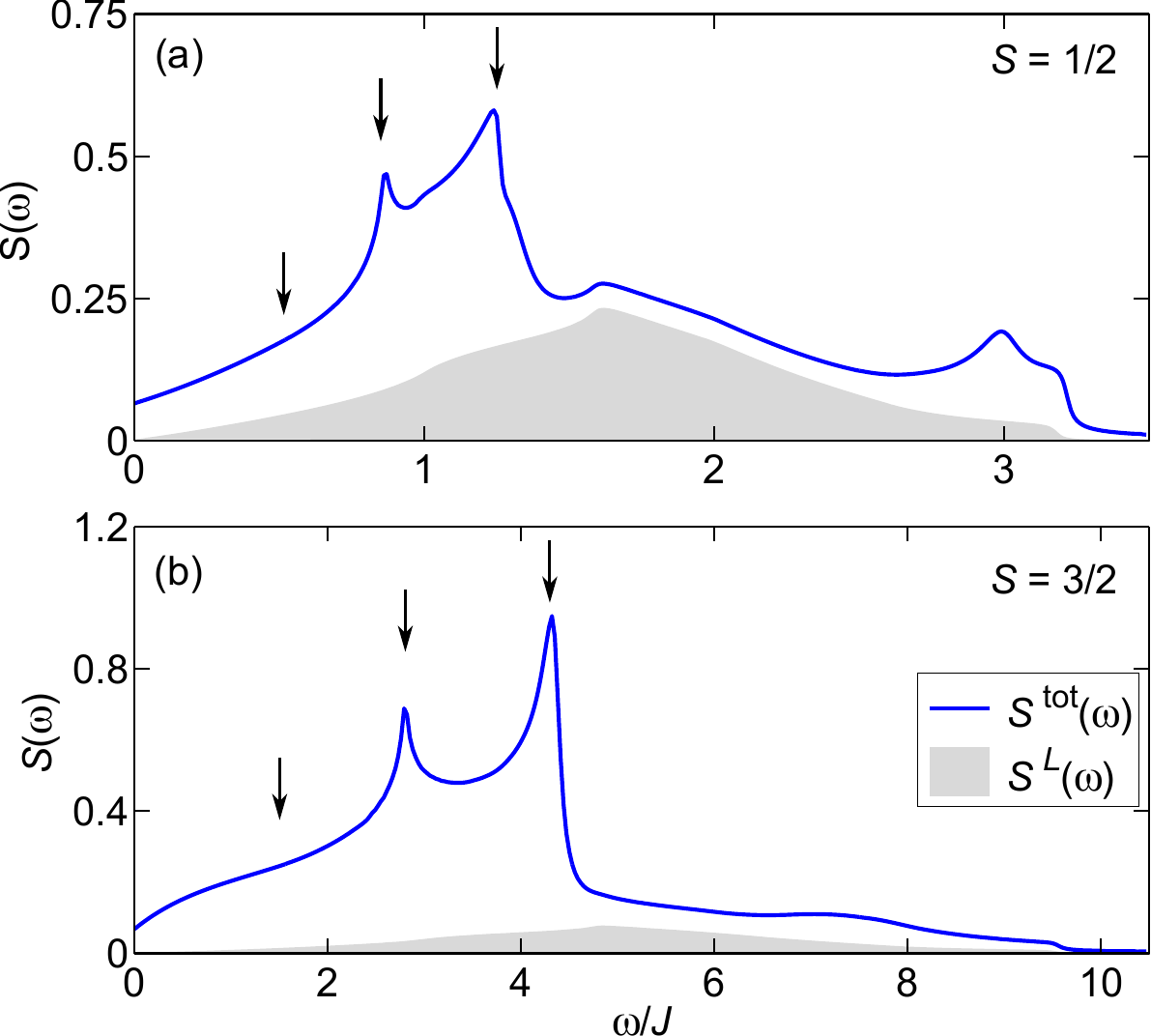}
\caption{(Color online) Energy dependence of the momentum-integrated dynamical structure factor ${\cal S}(\omega)$ for (a) $S=1/2$ and (b) $S=3/2$. Solid line corresponds to the total momentum-integrated structure factor ${\cal S}^{\rm tot}(\omega)$,  shaded area is the longitudinal component ${\cal S}^{L}(\omega)$. The vertical arrows indicate the energies
of the constant-energy scans in Fig.~\ref{fig6} and \ref{fig6a}.
}
\label{fig5}
\end{figure}

Figure~\ref{fig5} shows ${\cal S}^{\rm tot}(\omega)$ (solid line) and its longitudinal  component ${\cal S}^{L}(\omega)$ (shaded area) for $S=1/2$ and $S=3/2$. Strong peaks are observed for both values of spin. They are clearly identifiable  as the van Hove singularities in the spectra of dispersive quasiparticle-like modes. The lowest peak is associated with the high density of states at the magnon dispersion minimum at the M-point ($\omega\approx 2JS$), which also retains a considerable flatness in the MX direction. The upper peak is the standard van Hove singularity due to the top of the single-magnon spectrum ($\omega\approx 3JS$), which also has only weak dispersion along certain directions and thus contributes significantly to the density of states. Although these features appear less pronounced in the momentum-resolved dynamical structure factor  in Fig.~\ref{fig4}  for $S=1/2$ compared to $S=3/2$, strong peaks in the integrated spectrum in Fig.~\ref{fig5}(a) are still present.  Their energies can serve to estimate exchange constants and excitation bandwidth, e.g., from the powder-averaged neutron scattering experiments. We also note that no sign of the ``flat  band''  feature, advocated in Ref.~\onlinecite{Starykh06} for the renormalized on-shell spectrum of the HTAF in the $S=1/2$ case, is observed in ${\cal S}^{\rm tot}(\omega)$ in Fig.~\ref{fig5}(a).

What is most remarkable in the integrated structure factor in Fig.~\ref{fig5}, is that a significant spectral weight extends far beyond the upper edge of the single-magnon spectrum, the latter identifiable by the van Hove singularity. This behavior is not unlike the one recently observed in an $S=3/2$ triangular-lattice antiferromagnet.\cite{Toth12} It also highlights, once more, the necessity of taking magnon-magnon interaction into account in going beyond  the predictions of the linear SWT for the dynamical structure factor of non-collinear antiferromagnets. For the $S=1/2$ case, Fig.~\ref{fig5}(a), a massive contribution of the longitudinal ${\cal S}^{L}(\omega)$ to the high-energy spectral-weight is also rather spectacular.

\begin{figure}[th!]
\includegraphics[width=0.75\columnwidth]{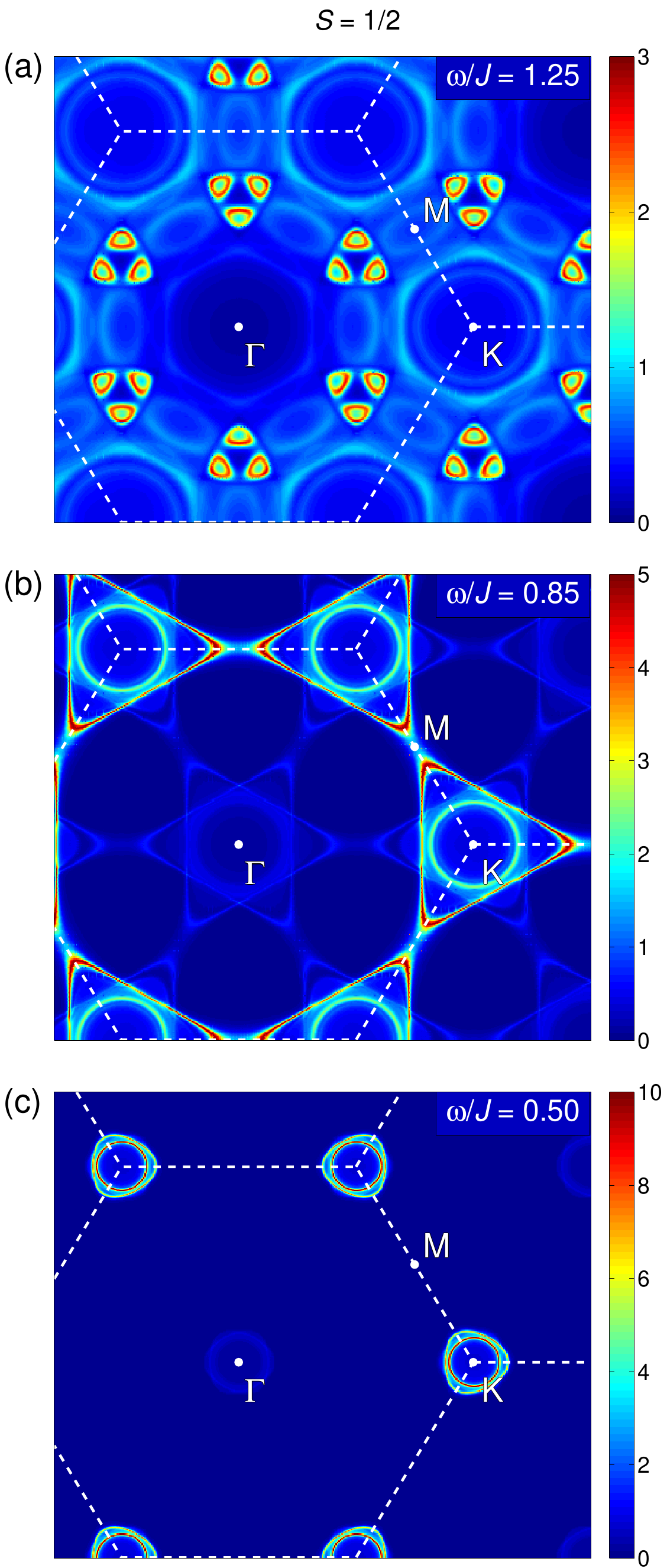}
\caption{(Color online) Intensity plots of the
constant-$\omega$ scans of the dynamical structure factor $S^{\rm tot}({\bf q,\omega})$ in the ${\bf q}$-plane
for $S=1/2$.  (a) $\omega/J=$1.25, (b) 0.85, and (c) 0.5.
Energies are indicated in ${\cal S}^{\rm tot}(\omega)$ in Fig.~\ref{fig5}(a).}
\label{fig6}
\end{figure}
\begin{figure}[th!]
\includegraphics[width=0.75\columnwidth]{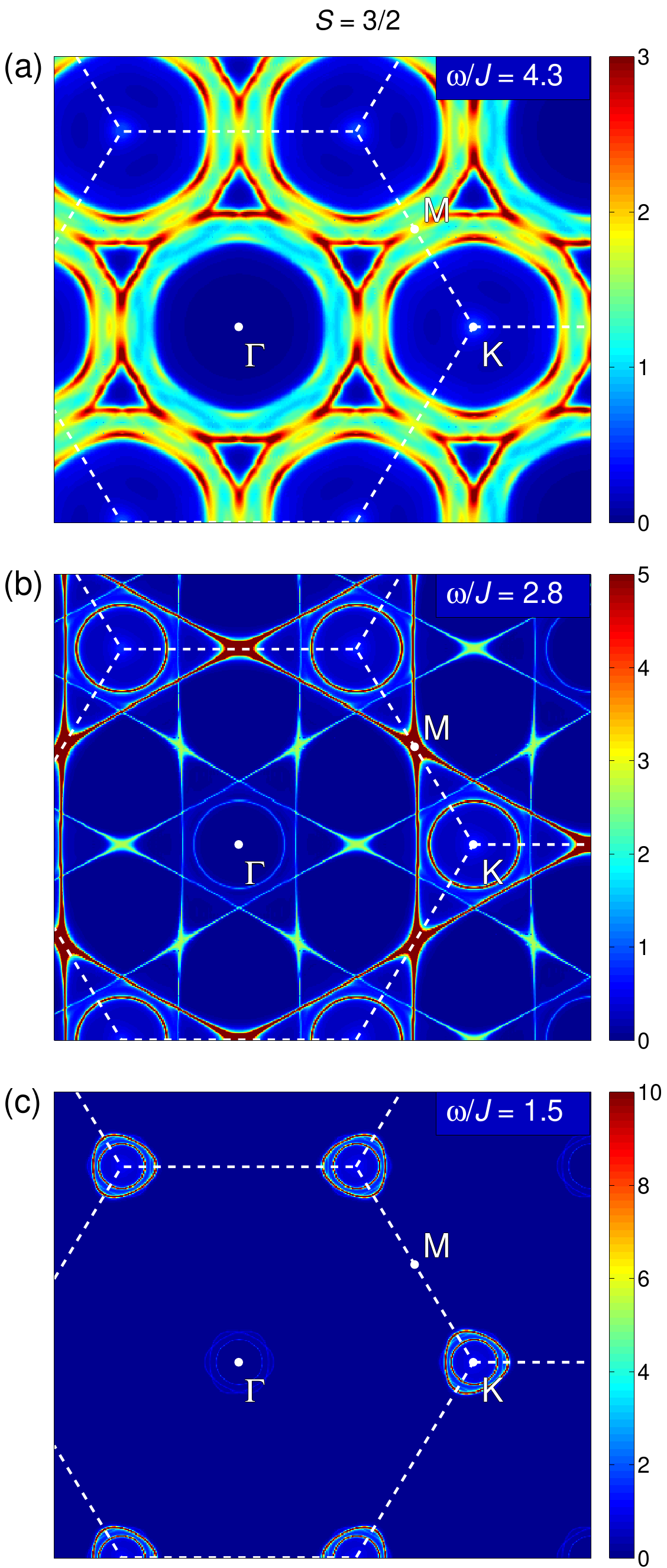}
\caption{(Color online) Same as in Fig.~\ref{fig6}
for $S=3/2$.  (a) $\omega/J=$4.3, (b) 2.8, and (c) 1.5.
Energies are indicated in ${\cal S}^{\rm tot}(\omega)$ in Fig.~\ref{fig5}(b).}
\label{fig6a}
\end{figure}

Complimentary to both Figs.~\ref{fig4} and \ref{fig5}, in Fig.~\ref{fig6} and \ref{fig6a} we present the constant-energy scans of the dynamical structure factor ${\cal S}^{\rm tot}({\bf q},\omega)$ for three selected energies (indicated as vertical arrows in Fig.~\ref{fig5}). Modern neutron scattering instrumentation is naturally suited for the studies of the dynamical correlations  in large regions of momentum space at fixed energies, which also motivates such a representation. One of the advantages of such constant-$\omega$ scans is that well-defined spin-wave excitations and corresponding van Hove singularities appear as bright sharp lines that are easy to distinguish from continuum scattering, manifested as broadly distributed diffuse intensity. The results for $S=1/2$ and $S=3/2$ are discussed below.

Figures~\ref{fig6}(a) and \ref{fig6a}(a) show the spectral weight near the top of the single-particle
spectrum, $\omega/J = 1.25$ and 4.3 for $S=1/2$ and $S=3/2$, respectively. In Fig.~\ref{fig6}(a), a well-defined spectrum is observed only  in the vicinity of two-thirds of the $\Gamma$M-line,  corresponding to excitations outside of the decay region, see also Fig.~\ref{fig4}. This should be compared to Fig.~\ref{fig6a}(a) for the $S=3/2$ case, exhibiting strong signal from almost flat branches of well-defined excitations contributing to the strong van Hove singularity observed in ${\cal S}^{\rm tot}(\omega)$ in Fig.~\ref{fig5}(b). The rest of the Brillouin zone in Fig.~\ref{fig6}(a) displays a weaker diffuse continuum, originating from broadened quasiparticle peaks and two-magnon continuum. Previously discussed features, such as  ``blow-out'' around Y$_1$ and M points are also clearly visible.

The energy $\omega/J = 0.85$ in Fig.~\ref{fig6}(b) corresponds to the vicinity of the first peak in ${\cal S}^{\rm tot}(\omega)$ in Fig.~\ref{fig5}(a), associated with the roton-like minima at the M-points with almost flat dispersion along the MX-line, see Fig.~\ref{fig4}(a), the latter indicating well-defined magnon excitations via a triangular-shaped intensities.  Circular distribution of spectral weight around the K-points corresponds to magnons clearly broadened by decays. Fainter, diffuse-like contributions are also seen around $\Gamma$-point. This should be compared to $S=3/2$ case in Fig.~\ref{fig6a}(b), where saddle-point features of magnon dispersion are much sharper.

Finally, a representative cut in the acoustic regime of the spectrum is shown in Figs.~\ref{fig6}(c) and \ref{fig6a}(c). At these energies, the effect of decays is weaker and concentric distributions of spectral weight around the  K-point reveal three distinct acoustic spin-wave branches from superposition of various ${\bf q}$, ${\bf q}+{\bf Q}$ and ${\bf q}-{\bf Q}$ contributions. The innermost (circular) distribution of spectral weight corresponds to ${\bf q}\rightarrow\Gamma$ excitations while the outermost (rounded triangular) contributions are associated with spin-waves from ${\bf q}\rightarrow{\rm K, K'}$. Compared to $S=3/2$, spectral intensity in the $S=1/2$ case clearly retains some diffuse component.  Vicinity of the $\Gamma$-point hosts similar pattern, albeit strongly suppressed  by the ${\bf q}$-dependent factors.

\section{Conclusion}
\label{sec:con}

We have developed an analytical theory for the dynamical structure factor of the triangular-lattice 
Heisenberg  antiferromagnet and presented explicit numerical results for $S^{\alpha\alpha}({\bf q},\omega)$ in the case of $S=1/2$ and $S=3/2$. Our treatment includes  comparison of  different contributions to the dynamical structure factor at the $1/S$-order, ensures the correct form of the Green's functions at low-energy and uses a pseudo-on-shell approach to avoid spurious manifestations of an unphysical pole in the spectral function.  In this way, we determined the dominant contributions to the dynamical structure factor to facilitate a thorough computation of the excitation spectrum in the entire momentum-energy space. In particular, our analysis demonstrates that contributions from anomalous Green's functions and mixed transverse-longitudinal terms can be neglected. The obtained energy-dependence of the dynamical structure factor displays a rich interplay of quasiparticle- and continuum-like features. Although our analysis is purely two-dimensional, we anticipate that further softening of unphysical singularities in the energy-dependence can be achieved by increasing dimensionality such that our conclusions should remain valid for a wide range of realistic materials.

The role of magnon-magnon interactions and presence of decays is demonstrated through the energy- and momentum-resolved spectrum as well as the momentum integrated structure factor.  A multitude of complex phenomena are observable in both. This includes non-Lorentzian lineshapes, quasiparticle blowout, 
roton-like minima, as well as an overall downward renormalization, a rich density of states and contributions from van Hove singularities.  Both spectrum presentations, integrated and momentum-resolved, highlight that quantum fluctuations transfer significant spectral weight to the two-magnon continuum, visible 
in both the transverse and the longitudinal components of the dynamical structure factor, with the latter contributing strongly to the overall dynamical 
response for $S=1/2$.

Our results provide the first determination of the full dynamical structure factor for the isotropic HTAF within the framework of non-linear spin-wave theory. They are consistent with and go beyond prior studies on quasi-one-dimensional spiral and spatially anisotropic triangular antiferromagnets by maintaining proper treatment of Green's functions and achieving systematic ranking of different $1/S$ contributions. These detailed calculations provide a guide for experimental observation of the effects of magnon interaction and decays as well as a direct analytical scheme to predict the spin dynamics for realistic single-crystalline materials. Moreover, the inclusion of the momentum-integrated dynamical structure factor provides a guide for observations in materials  for which only powder samples are available.  Thus, this work presents the full landscape of the non-linear spin-wave dynamics in the triangular lattice Heisenberg antiferromagnet.

\begin{acknowledgments}
This work was supported in part by the US Department of Energy under grants  DE-FG02-08ER46544 (M.~M. and W.~T.~F.) and DE-FG02-04ER46174 (A.~L.~C.).
\end{acknowledgments}

\appendix
\section{Subleading corrections to the structure factor and derivation of ${\cal S}^{zz}$}
\label{app:A}
\subsection{Anomalous terms in the transverse structure factor}

Here we evaluate the contribution from the anomalous Green's function $G_{12}({\bf k},\omega)$ to the transverse structure factor. The Green's functions can be expressed explicitly from (\ref{BE}) as
\begin{eqnarray}
G_{11}({\bf q},\omega) &=&
	  \frac{1}{
		[ \omega-\varepsilon_{\bf q}-\Sigma_{11}({\bf q},\omega)] } \ , \\
G_{12}({\bf q},\omega) &=&
	  \frac{-\Sigma_{12}({\bf q},\omega)}{
		[ \omega-\varepsilon_{\bf q}-\Sigma_{11}({\bf q},\omega)]
		[ \omega+\varepsilon_{\bf q}+\Sigma_{11}(-{\bf q},-\omega)]} \ . \nonumber
\end{eqnarray}

\begin{figure}[t!]
\includegraphics[width=0.95\columnwidth]{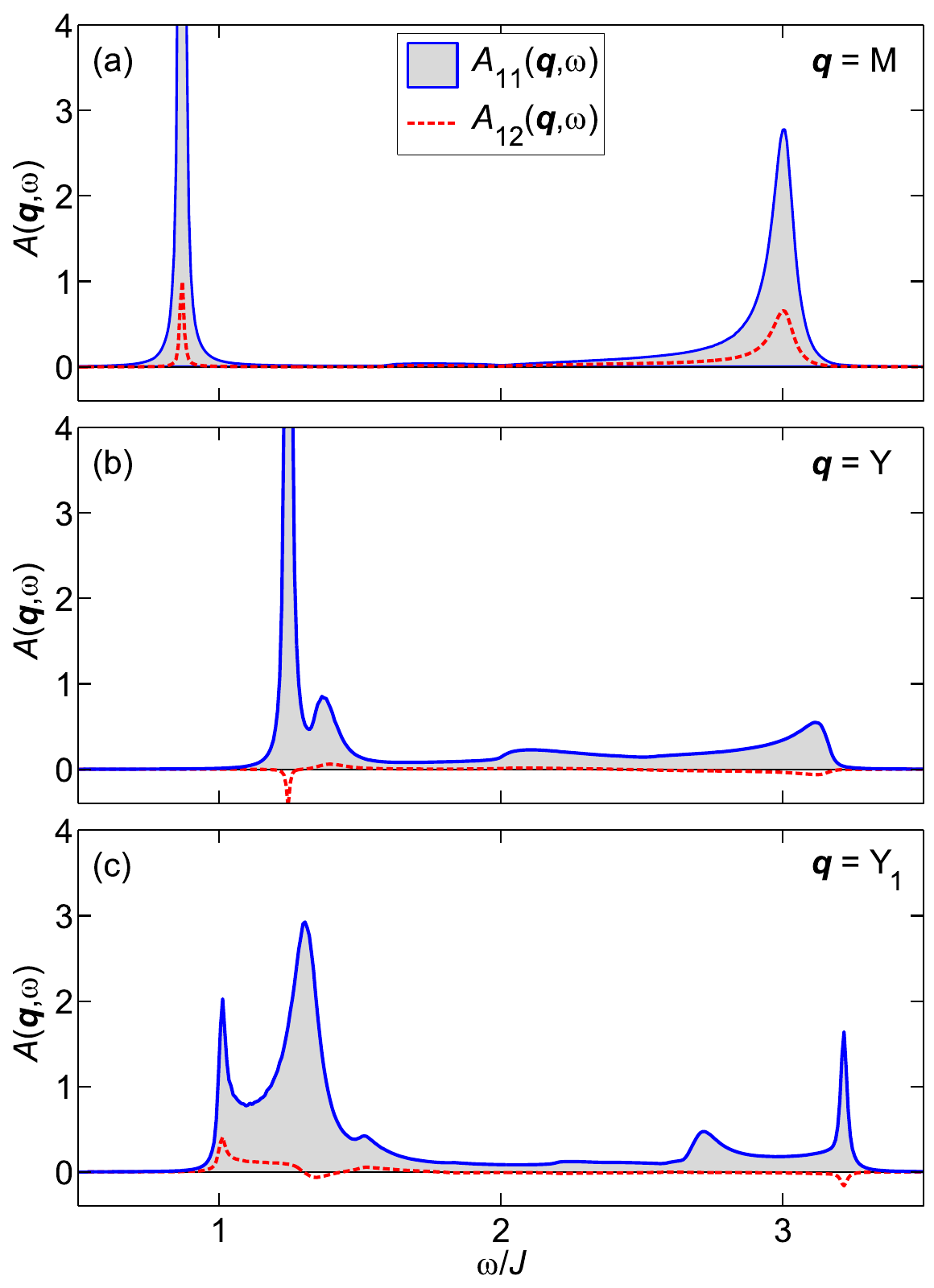}
\caption{(Color online) (a)--(c) Energy dependence of the contributions to the spectral function for $S=1/2$ at the M, Y and Y$_1$ {\bf q}-points, see Fig.~\ref{fig1}(b). The solid lines with shaded area correspond to $A_{11}({\bf q},\omega)$ and the lower dashed lines to $A_{12}({\bf q},\omega)$. The contributions from $A_{11}(-{\bf q},-\omega)$ are vanishingly small.}
\label{fig7}
\end{figure}

As is discussed in Sec.~\ref{sec:ResTF}, the unphysical mode needs to be controlled by keeping the source term in \eqref{Sigma} on-shell. This form of the self-energies is used in all our numerical evaluations. The relative role of normal and anomalous Green's function is estimated through their respective spectral functions,
\begin{equation}
A_{11,12}({\bf q},\omega) = -\frac{1}{\pi} {\rm Im}[G_{11,12}({\bf q},\omega)]
\ ,
\end{equation}
plotted  for representative high-symmetry points  in Fig.~\ref{fig7}. The dominant contribution comes from $A_{11}({\bf q},\omega)$ while $A_{12}({\bf q},\omega)$ yields much smaller contribution already for the $S=1/2$. In addition, its contributions are also redundant to that of  $A_{11}({\bf q},\omega)$ in terms of qualitative features. Overall, we conclude that the contribution from $A_{12}({\bf q},\omega)$ is small and that from $A_{11}(-{\bf q},-\omega)$ is zero away from the K-point, so that they both can be  neglected in (\ref{G_trans}).

\subsection{Longitudinal fluctuations}

We begin the derivation of ${\cal S}^{zz}({\bf q},\omega)$ with the expression for the longitudinal spin fluctuations in (\ref{dSz}). At zero temperature, ${\cal S}^{zz}({\bf q},\omega)$ is related to the time-ordered Green's function by
\begin{eqnarray}
&& {\cal G}^{zz}({\bf q},t) = -i \left\langle T \delta S_{\bf q}^z(t)\delta S_{-{\bf q}}^z\right\rangle
\nonumber \\
&& \phantom{{\cal G}^{zz}} =
 -\frac{i}{N} \sum_{{\bf k}_1,{\bf k}_2} \bigl\langle Ta_{{\bf k}_1}^\dagger(t) a_{{\bf k}_1+{\bf q}}(t)
a_{{\bf k}_2}^\dagger a_{{\bf k}_2-{\bf q}}\rangle .
\label{G_long}
\end{eqnarray}
This correlator probes the two-particle density of states and thus provides information about the two-magnon continuum.

The longitudinal component of the structure factor is of the order of $O(1/S)$, a factor of $1/S$ smaller than the leading terms in the transverse correlation function (\ref{G_trans}). Then, in the spirit of the $ 1/S$ expansion, ${\cal G}^{zz}$ in (\ref{G_long}) can be calculated without taking into account interaction corrections.

Performing Bogolyubov transformation in  (\ref{G_long})  and keeping  terms only with two creation and  two annihilation operators we obtain
\begin{eqnarray}
&& {\cal G}^{zz}({\bf q},t)\! = \!
i\! \sum_{\bf k}(u_{\bf k} u_{\bf k+q}\! + v_{\bf k} v_{\bf k+q})^2 G({\bf k},\!-t)G({\bf k\!+\!q},t)
\nonumber \\
&& \mbox{} +   \frac{i}{2}\! \sum_{\bf k}
\bigl(u_{\bf k} v_{\bf k+q} + v_{\bf k} u_{\bf k+q} \bigr)^2
G({\bf k},-t)G({\bf -k-q},-t)
\nonumber \\
& & \mbox{} +
\frac{i}{2}\!\sum_{\bf k}
(u_{\bf k} v_{\bf k-q}\! + v_{\bf k} u_{\bf k-q} )^2 G({\bf k},t)G({\bf -k+q},t),
\label{Gzz}
\end{eqnarray}
where $G=G_{11}$ for brevity. Subsequent transformation to $\omega$-representation shows that the first two terms in (\ref{Gzz}) have no imaginary part in the  noninteracting  approximation. Hence, the leading contribution to the longitudinal structure factor is given by the last term:
\begin{eqnarray}
&& {\cal S}^{zz} ({\bf q},\omega) = \frac{1}{2\pi}
\sum_{\bf k} (u_{\bf k} v_{\bf k-q} + v_{\bf k} u_{\bf k-q})^2 \nonumber \\
&& \phantom{{\cal S}^{zz}({\bf q}} \times \textrm{Im}\!\int \frac{d\omega'}{2\pi i}\,
G({\bf k},\omega') G({\bf -k+q},\omega-\omega').
\label{Szz}
\end{eqnarray}
The remaining integral can be taken for $G({\bf k},\omega)=G^0_{11}({\bf k},\omega)$ to yield Eq.~(\ref{Szzqw}).

To improve upon the linear SWT approximation, one can  use  (\ref{Szz}) with the interacting  Green's functions. In such an approximation the single-magnon energies will be renormalized by interactions, while still neglecting other effects of interactions in the correlation function.

\subsection{Mixed transverse-longitudinal fluctuations}

Here we provide a few remarks concerning the mixed transverse-longitudinal correlators. We use the identity
\begin{eqnarray}
	i\bigl[ {\cal S}^{xz}({\bf q},\omega) &-& {\cal S}^{zx}({\bf q},\omega)\bigr]
\label{eq:Gxz}\\
& =& -\frac{1}{\pi}\, \textrm{Im}\,\Big\{{\cal G}^{xz} ({\bf q},\omega) - {\cal G}^{zx} ({\bf q},\omega)\Big\}\ ,
\nonumber
\end{eqnarray}
where the corresponding Green's functions are defined as
\begin{eqnarray}
&&{\cal G}^{xz} ({\bf q},t)  = \left\langle TS^x_{\bf q}(t)S^z_{-\bf q}\right\rangle,\nonumber\\
&&{\cal G}^{zx} ({\bf q},t)  = \left\langle TS^z_{\bf q}(t)S^x_{-\bf q}\right\rangle.
\label{Gxz0}
\end{eqnarray}
The above identity is derived by applying the fluctuation dissipation theorem (\ref{FDT0}) to time-dependent fluctuations
of the operator $\hat{A}= S^x_{\bf q} - i S^z_{\bf q}$ (with $\hat{A}^\dagger = S^x_{-\bf q} + i S^z_{-\bf q}$) and
excluding parts that are  diagonal in spin indices, which have been considered before.
Using bosonic representation for spin operators we obtain
\begin{eqnarray}
{\cal G}^{xz}({\bf q},t)&=&-(1-\Lambda_+)\sqrt{\frac{S}{2N}}
\label{Gxz00}\\
&&\times\sum_{\bf k}
\langle T \bigl[a_{\bf q}(t)+a^\dagger_{-\bf q}(t)\bigr]a^\dagger_{\bf k} a_{\bf k-q}\rangle \ , \nonumber
\end{eqnarray}
and a similar expression for ${\cal G}^{xz}({\bf q},t)$.

\begin{figure}[t!]
\includegraphics[width=0.9999\columnwidth]{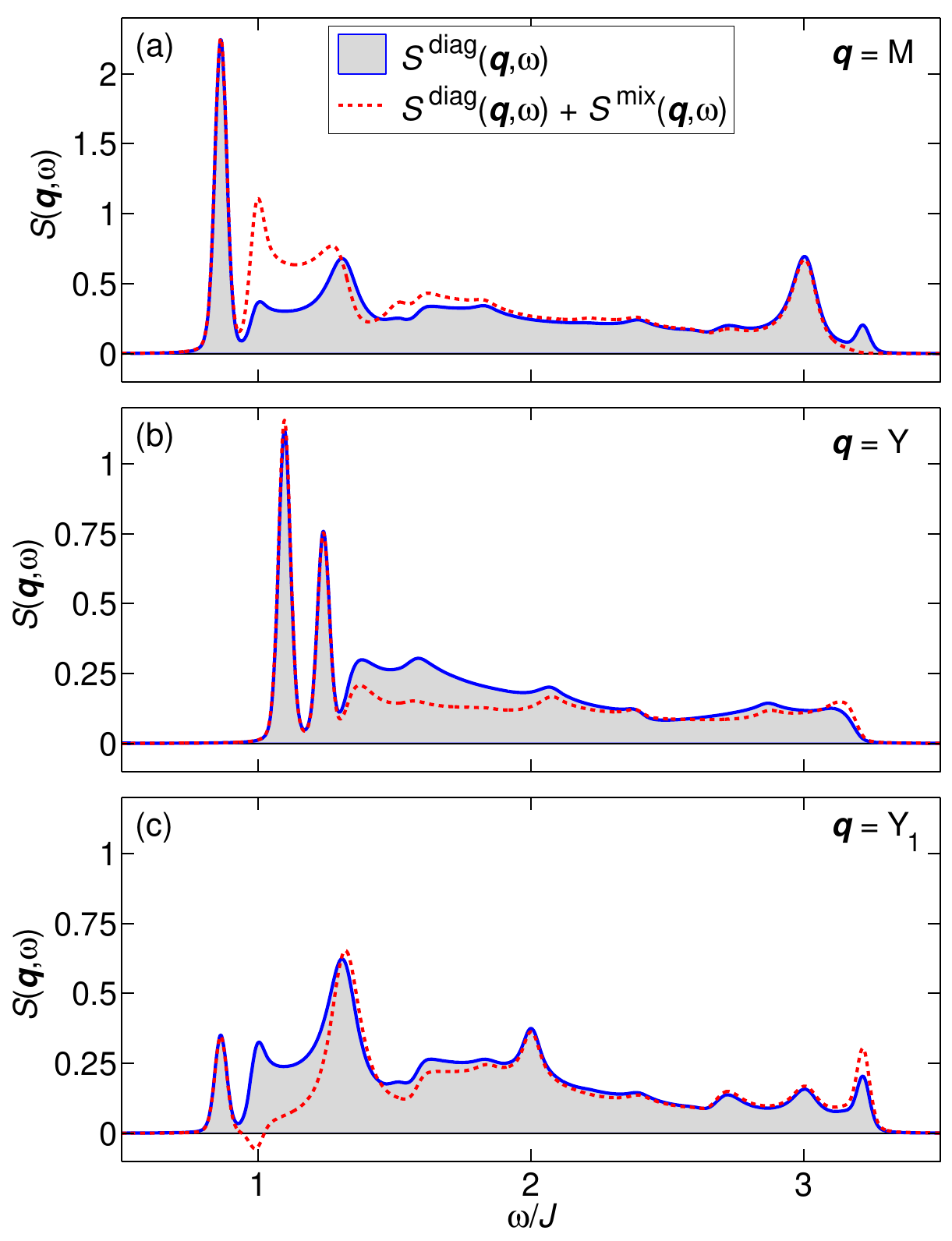}
\caption{(Color online) (a)--(c) Energy dependence of the total dynamical structure factor ${\cal S}^{\rm tot}({\bf q},\omega) = {\cal S}^{\rm diag}({\bf q},\omega) + {\cal S}^{\rm mix}({\bf q},\omega)$, Eq.~(\ref{Stot_1}), dashed lines, compared to the diagonal part (solid lines with shaded areas) for $S=1/2$ at representative  ${\bf q}$-points M, Y and Y$_1$. The results are convoluted with a Gaussian profile of width $\sigma = 0.03J$ as in Fig.~\ref{fig3}. }
\label{fig8}
\end{figure}

The first nonzero contribution to the mixed Green's functions comes from the first-order
perturbation term in three-particle interaction $\hat{V}_3$. This means that the mixed
correlator gives a $O(1/S)$ contribution compared to the transverse structure factor (\ref{G_trans})
and is, formally, of the same $1/S$ order as ${\cal S}^{zz}$ (\ref{Szzqw}). Keeping
terms that are nonzero in the noninteracting limit
and performing standard calculations we obtain
\begin{eqnarray}
&& {\cal G}^{xz} ({\bf q},\omega)  =
-\frac{3S\sqrt{3}}{4}(1-\Lambda_+)(u_{\bf q} + v_{\bf q})\nonumber\\
&&\phantom{{\cal G}^{xz} ({\bf q},\omega)  =} \times\sum_{\bf k}(u_{\bf k} v_{\bf q-k} + v_{\bf k} u_{\bf q-k})
 \nonumber \\
&& \times    \biggl\{ G({\bf q},\omega)
\Bigl[\frac{\tilde{V}_{31}({\bf k};{\bf q})}{\omega-\varepsilon_{\bf k} - \varepsilon_{\bf q-k}+i0}
+ \frac{\tilde{V}_{32}(-{\bf k},{\bf q})}{\omega+\varepsilon_{\bf k} + \varepsilon_{\bf q-k}}\Bigr]
\label{Gxz11} \\
&& + G(-{\bf q},-\omega)\Bigl[\frac{\tilde{V}_{32}({\bf k},-{\bf q})}{\omega-\varepsilon_{\bf k} - \varepsilon_{\bf q-k}+i0}
+ \frac{\tilde{V}_{31}(-{\bf k};-{\bf q})}{\omega+\varepsilon_{\bf k} + \varepsilon_{\bf q-k}}\Bigr]\biggr\}\,,
\nonumber
\end{eqnarray}
while ${\cal G}^{zx} ({\bf q},\omega) = -{\cal G}^{xz} ({\bf q},\omega)$. In Eq.~\eqref{Gxz11}, $\tilde{V}_{31}({\bf k};{\bf q})$ and $\tilde{V}_{32}({\bf k},{\bf q})$ are the dimensionless cubic vertices from Ref.~\onlinecite{Chernyshev09}.
The final expression for the mixed dynamical structure factor of Eq.~(\ref{Stot_1}) reads
\begin{equation}
{\cal S}^{\rm mix}({\bf q},\omega) = i\Bigl[{\cal S}^{xz}({\bf q-Q},\omega) - {\cal S}^{xz}({\bf q+Q},\omega) \Bigr] ,
\end{equation}
where $i{\cal S}^{xz}({\bf q},\omega)=-(1/\pi)\textrm{Im}\bigl[{\cal G}^{xz}({\bf q},\omega)\bigr]$ with ${\cal G}^{xz}$
from (\ref{Gxz11}). For the calculations presented in Fig.~\ref{fig8}, the Green's functions are
taken as $G({\bf q},\omega)=G_{11}({\bf q},\omega)$ from Eq.~(\ref{BE}) with
$\Sigma_{11}({\bf q},\omega)$ given by Eq.~\eqref{Sigma}.
 In this approximation, the single- and the two-magnon contributions to the mixed structure
factor are evaluated with the same accuracy as in the transverse and the longitudinal components
${\cal S}^{xx}$, ${\cal S}^{yy}$ and ${\cal S}^{zz}$ in (\ref{G_trans_1}) and (\ref{Szzqw}),
respectively. Because of the nature of our calculations, higher-order  $1/S$ terms are not selfconsistently accounted for,
leading to an unphysical overcompensation in Fig.~\ref{fig8}(c), where a small portion of ${\cal S}^{\rm tot}({\bf q},\omega)$
is slightly below zero, presumably an $O(1/S^2)$ effect.

Because of the  contributions having opposite signs in (\ref{Gxz11}), it can be anticipated that ${\cal S}^{\rm mix}$yields a subleading correction to  the transverse  ${\cal S}^{\perp}$ in (\ref{G_trans}) and to the two-magnon continuum ${\cal S}^{L}$ in (\ref{Szzqw}) despite ${\cal S}^{zz}$ and ${\cal S}^{xz}$ being formally of the same $1/S$-order. This is confirmed in Fig.~\ref{fig8}, where Eq.~(\ref{Stot_1}) is used to evaluate ${\cal S}^{\rm diag}$ and ${\cal S}^{\rm mix}$ in the total dynamical structure factor ${\cal S}^{\rm tot} = {\cal S}^{\rm diag}  + {\cal S}^{\rm mix}$.

First, the impact of ${\cal S}^{\rm mix}$  on the dominant peaks  in ${\cal S}^{\rm diag}$ is vanishingly small, showing that the mixed terms do not affect the leading quasiparticle-like part of the spectrum.  Then, we observe that some of the edge-singularities in ${\cal S}^{\rm tot}$ are enhanced while some are suppressed by the inclusion of the mixed term, suggesting that the latter would not yield an overall regularization of such singularities. Primary effect of  ${\cal S}^{\rm mix}$ is a  modulation of a relatively small part  of the broad two-magnon continuum in ${\cal S}^{\rm diag}$.

Most importantly, the contribution from ${\cal S}^{\rm mix}({\bf q},\omega)$ to the momentum-integrated structure factor  ${\cal S}(\omega)$ is {\it exactly} zero due to the antisymmetric properties of ${\cal S}^{xz}({\bf q},\omega)$ discussed above. Therefore, we conclude that the off-diagonal term ${\cal S}^{\rm mix}$ can be neglected compared to the leading transverse ${\cal S}^{\perp}$ and longitudinal ${\cal S}^{L}$  terms in ${\cal S}^{\rm diag}$. This strongly justifies the choice  ${\cal S}^{\rm tot} \approx {\cal S}^{\rm diag}$ used in Sec.~\ref{sec:DC} and Sec.~\ref{sec:Res}.
\clearpage



\end{document}